\documentclass[journal]{IEEEtran}
 \usepackage{amsmath,amssymb}
 \usepackage{subfigure}
 \usepackage{graphicx,graphics,color,psfrag}
 \usepackage{cite,balance}
 \usepackage{caption}
 \allowdisplaybreaks
 \usepackage{algorithm}
 \usepackage{accents}
 \usepackage{amsthm}
 \usepackage{bm}
 \usepackage{algorithmic}
 \usepackage[english]{babel}
 \usepackage{multirow}
 \usepackage{enumerate}
 \usepackage{cases}
 \usepackage{stfloats}
 \usepackage{dsfont}
 \usepackage{color,soul}
 \usepackage{amsfonts}
 \usepackage{cite,graphicx,amsmath,amssymb}
 \usepackage{subfigure}
 \usepackage{fancyhdr}
 \usepackage{hhline}
 \usepackage{graphicx,graphics}
 \usepackage{array,color}
 \usepackage{amsmath}
 \usepackage{booktabs}
\include{header}

\begin{document}
\setlength{\topskip}{-3pt}

\graphicspath{{Figs/}}

\title{Performance Analysis for Heterogeneous Air- Ground ISAC in Coordinated Multipoint Networks}
\author{Yihang Jiang, Xiaoyang Li, Guangxu Zhu, Changsheng You, Xiaowen Cao, Dingzhu Wen, Bingpeng Zhou, Xinyi Wang, and Rui Zhang\\
\thanks{This work is supported in part by the National Natural Science Foundation of China under Grants U25A20389, U25A20394, 62371478, 62501407, 62571227, 62331023, and 62401369, Young Elite Scientists Sponsorship Program by CAST under Grant YESS20240364, Guangdong Major Project of  Basic and Applied Basic Research under Grant 2023B0303000001, Guangdong Science and Technology Program under Grants 2026A1515010226, 2024A1515010097, 2026A1515012153, and 2024A1515012259, Guangdong Young Talent Research Project under Grant 2023TQ07A708, Shenzhen Science and Technology Program under Grants KJZD20240903095402004, JCYJ20241202124934046, JCYJ20240813094212016, and JCYJ20240813151253068, Guangdong Provincial Key Laboratory of Future Networks of Intelligence under Grant 2022B1212010001, Shenzhen Loop Area Institute under Grant SLAI2026020007, Science and Technology Commission of Shanghai Municipality under Grant 25DP1501900, the Program under Grant 2023QN10X152, and Shanghai Sailing Program under Grant 23YF1427400. (Corresponding author: Xiaoyang Li)

Yihang Jiang and Guangxu Zhu are with the Shenzhen Research Institute of Big Data, The Chinese University of Hong Kong-Shenzhen, Shenzhen, 518172, China. (Email: yihangjiang1@link.cuhk.edu.cn, gxzhu@sribd.cn)

Xiaoyang Li and Changsheng You are with the Department of Electrical and Electronic Engineering, Southern University of Science and Technology, Shenzhen, 518055, China. (Email: lixy@sustech.edu.cn, youcs@sustech.edu.cn) 

Xiaowen Cao is with the College of Electronic and Information Engineering, Shenzhen University, Shenzhen, 518100, and with Guangdong Provincial Key Laboratory of Future Networks of Intelligence, Shenzhen, 518172, China. (email: caoxwen@szu.edu.cn )

Dingzhu Wen is with the School of Information Science and Technology, ShanghaiTech University, Shanghai, 201210, China. (Email: wendzh@shanghaitech.edu.cn)  

Bingpeng Zhou is with the School of Electronics and Communication Engineering, Sun Yat-sen University, Shenzhen, 518000, China. (Email: zhoubp3@mail.sysu.edu.cn)  

Xinyi Wang is with the School of Information and Electronics, Beijing Institute of Technology, Beijing, 100081, China. (Email: wangxinyi@bit.edu.cn)  

Rui Zhang is with the Department of Electrical and Computer Engineering, National University of Singapore, Singapore. (Email: elezhang@nus.edu.sg)}
}

\maketitle

\begin{abstract}
The emergence of the \textit{low-altitude economy} (LAE) calls for highly integrated and reliable wireless systems that can simultaneously support \textit{communication and sensing} (C\&S) functions. Although \textit{integrated sensing and communication} (ISAC) has been widely studied, most existing works focused on link-level or single-cell architectures in terrestrial environments, leaving the potential of network-level cooperative air-ground ISAC largely unexplored. To bridge this gap, a heterogeneous air-ground ISAC network architecture based on \textit{coordinated multipoint} (CoMP) is proposed, which incorporates a cooperative hybrid mono/bi-static sensing scheme to enhance spatial diversity and sensing capability. In the proposed architecture, a two-tier \textit{base station} (BS) deployment is adopted: master BSs are arranged in a hexagonal lattice, while slave BSs follow a Poisson point process distribution. This structure concurrently supports communication for terrestrial users and sensing for aerial targets. A holistic performance analysis framework for both C\&S is further developed, accounting for key channel and network parameters. Simulation results reveal inherent trade-offs between C\&S performance, especially under multi-BS cooperation and varying network density. These findings provide practical guidance for the deployment of scalable and efficient ISAC networks in LAE scenarios.
\end{abstract}

\newpage


\section{Introduction}
The \textit{low-altitude economy} (LAE) has emerged as a promising concept that utilizes low-altitude airspace to enable innovative aerial services with substantial socioeconomic impacts \cite{jiang2025integrated}. The sustainable development of this emerging sector necessitates robust safety assurance mechanisms, particularly through ubiquitous communication infrastructures and real-time sensing networks capable of supporting reliable, high-density aerial operations. These technological requirements create a natural synergy with \textit{integrated sensing and communication} (ISAC) systems \cite{fei2023air, jiang2022design, jiang2023design}, which have gained prominence as a key enabling technology for LAE applications.

As a fundamental technological solution for LAE, ISAC achieves deep integration of traditionally separated radar sensing and wireless communication functions within a unified system architecture. This convergence delivers substantial performance improvements through shared spectrum resources, hardware platforms, and signal processing capabilities \cite{liu2022integrated,xia2025symbiotic}. The existing ISAC research has predominantly addressed link-level and system-level optimization, with particular emphasis on waveform design and resource management within single \textit{base station} (BS) architectures. Nevertheless, the broader potential of network-level ISAC, especially multi-cell cooperative \textit{communication and sensing} (C\&S) paradigms, has not been sufficiently explored in existing literature \cite{meng2024network,jiang2025network}.

As LAE activities raise urgent requirements of wide-area C\&S, such gap in system modeling and performance analysis of network-level ISAC becomes critical. Thorough analysis would lay a solid foundation for network optimization to meet diverse operational requirements, including safety, reliability, and scalability.

\subsection{Related Work}
As a powerful mathematical tool, \textit{stochastic geometry} (SG) \cite{chiu2013stochastic, haenggi2012stochastic} has been widely adopted for analyzing C\&S networks. Prior studies have systematically investigated ISAC network performance under diverse operational conditions. For instance, \cite{ren2018performance} quantitatively examined interference effects in time-division ISAC systems that separates C\&S functions into distinct intervals. Building upon this foundation, \cite{sun2024performance} accounted for the impact of urban building obstructions on the specified C\&S coverage probability. Meanwhile, \cite{olson2023coverage} developed a comprehensive analytical framework to evaluate joint C\&S capabilities, deriving closed-form expressions for both coverage probability and ergodic rate in ISAC networks. Additionally, \cite{salem2024rethinking} proposed a \textit{beamforming} (BF) design for ISAC networks, followed by an analysis of spectral and energy efficiency.

Although these pioneering studies have made significant progress in ISAC network modeling and performance analysis, most studies only considered individual cells, ignoring the potential advantages of multi-cell collaboration. In practice, multi-BS cooperation can substantially enhance ISAC performance through coordinated BF, distributed sensing data/information fusion, and network-level resource optimization. Recent studies have begun investigating cooperative multi-BS ISAC architectures to address this gap. The work by Meng et al. \cite{meng2023network} proposed a dual clustering strategy that jointly optimizes \textit{communication user} (CU)-centric and \textit{sensing target} (ST)-centric configurations, providing comprehensive analysis of area spectral efficiency while introducing an interference-aware cooperative BF scheme to mitigate inter-BS sensing interference. Similarly, Gan et al. \cite{gan2024coverage} developed a CU-centric multi-BS cooperative ISAC framework for joint downlink communication and CU self-positioning, enabling rigorous joint performance evaluation through coverage probability and ergodic rate metrics with sensing performance quantified via \textit{Cram\'er-Rao Bound} (CRB)-constrained coverage probability. 

While these studies demonstrated that collaborative architectures can provide additional \textit{degrees of freedom} (DoF) to flexibly optimize and balance C\&S performance, they remain fundamentally constrained by point-to-point C\&S services relying on nearest-neighbor or strongest-signal association strategies. Such approaches cannot fully exploit the potential collaborative gains of multi-cell systems. Advanced distributed \textit{multiple-input multiple-output} (MIMO) and \textit{coordinated multipoint} (CoMP) networks offer promising solutions, as they not only mitigate inter-cell interference but also enhance edge-user service quality through coordinated multi-point transmission \cite{hosseini2016stochastic}. However, research on ISAC network modeling and performance analysis within such architectures remains limited. A recent study \cite{meng2024cooperative} examined this problem by investigating cooperative BS cluster sizes for both CU and ST-centric scenarios to achieve balanced C\&S performance tradeoffs.

Despite their benefits, CoMP networks face significant implementation challenges, particularly regarding \textit{channel state information} (CSI) acquisition and centralized processing requirements. The dependence on pre-deployed fronthaul links makes dynamic adaptation of network topology according to CU/ST locations particularly difficult. This limitation makes localized collaboration strategies based on BS locations a more practical alternative to CU/ST-centric coordination for ISAC networks. 

Furthermore, existing studies on terrestrial networks have largely overlooked the critical influence of altitude on C\&S performance. To bridge this research gap, our recent work \cite{jiang2024coverage, jiang2025network} has pioneered the investigation of integrated air-ground ISAC networks. Specifically, \cite{jiang2025network} proposed a comprehensive analytical framework for evaluating network-level performance in cooperative air-ground ISAC networks. This framework enables quantitative assessment of C\&S performance with multiple metrics, including: (1) area \textit{communication coverage probability} (CCP), (2) area communication spectral efficiency, (3) area \textit{radar detection coverage probability} (RDCP) under a \textit{constant false alarm rate} (CFAR) criterion, and (4) \textit{average CRB} (ACRB). However, this work has not yet incorporated advanced CoMP architectures into the analysis. 

In addition, in the context of cooperative sensing, most existing work on multi-BS sensing has primarily considered a mono-static sensing architecture. Although this architecture can provide spatial diversity gain through multi-view observations, it fails to fully unlock the potential of multi-BS collaboration. In contrast, bi-static sensing has the potential to achieve shorter sensing link distances, thereby further enhancing cooperative sensing performance. Therefore, in this work, a cooperative hybrid mono/bi-static sensing architecture is considered, where mono-static sensing is retained to guarantee a performance lower bound for sensing tasks when shorter bi-static links are unavailable.

\subsection{Contributions}
To address the aforementioned limitations, a heterogeneous air-ground ISAC CoMP network is proposed in this paper, where \textit{orthogonal frequency-division multiplexing} (OFDM) signaling is employed to balance C\&S functionalities. Based on this network, a network-level performance analysis framework for C\&S performance evaluation is further established. The effects of key system parameters on C\&S performances are further analyzed and verified via simulations. The main contributions of this paper are summarized as follows.

\begin{itemize}
\item \textbf{\textit{Modeling and Analysis Framework for Heterogeneous Air-Ground ISAC CoMP Networks:}} Focusing on the need for key enablers in LAE applications, a comprehensive framework is proposed for modeling and analyzing a novel cooperative heterogeneous air-ground ISAC network based on CoMP transmission in this paper. The system consists of two tiers of BSs: The master BSs are deployed according to a conventional hexagonal lattice layout, while the slave BSs are distributed following a \textit{homogeneous Poisson point process} (HPPP) model. This structure simultaneously supports communication services for terrestrial users and sensing functions for aerial targets. In contrast to traditional terrestrial ISAC networks, the proposed model incorporates critical additional parameters such as altitude. Furthermore, a cooperative hybrid mono/bi-static sensing architecture is adopted to enhance spatial diversity and improve sensing performance.

\item \textbf{\textit{Statistical Modeling and Comprehensive Performance Analysis for ISAC Networks:}} This work establishes a systematic analytical framework by first deriving fundamental statistical characterizations, including the distributions of received signal power and link distances along with their Laplace transforms for both C\&S. Building upon this mathematical foundation, a comprehensive performance analysis is conducted: for communication, it employs both per-CU and per-BS metrics, specifically the CCP and the \textit{ergodic efficiency} (EE), with the latter analyzed across individual \textit{resource elements} (REs) and aggregated multi-carrier; for wireless sensing, the evaluation incorporates both link-level and network-level metrics, including the RDCP under a CFAR criterion and the ACRB for bi-static links, as well as the \textit{radar cumulative detection probability} (RCDP) for cooperative multi-link sensing.

\item \textbf{\textit{Network Deployment Guidelines:}} Extensive simulations validate our theoretical analysis and examine key parameter impacts. The results indicate that increasing the number of cooperative slave BSs $N_l$ can reduce the BF gain in communication interference signals while enhancing the coherent combining gain in communication effective signals, thereby improving per-RE communication performance. However, under multi-carrier aggregation conditions, performance degrades due to the reduced number of subcarriers assignable to each CU, revealing a trade-off between BF gains and spectrum allocation. Additionally, for traditional ground networks, the balance between changes in effective signals and interference signals caused by variations in BS density $\lambda_B$ renders its impact on C\&S performance negligible. It solely affects the air-ground network performance evaluated through per-EE and per-link metrics. These findings, along with additional detailed analysis of the loading factor $\eta_c$ and OFDM configurations (subcarrier amount $N$ and symbol amount $M$) offers practical guidance for deploying heterogeneous air-ground ISAC CoMP network with balanced C\&S performance.
\end{itemize}

The remainder of this paper is organized as follows. Section II introduces the system model for the heterogeneous air-ground ISAC CoMP network. Section III presents the analysis of distriutions, including the statistical distributions of signal power and link distance. The performance analysis for C\&S is conducted in Section IV and Section V, respectively, using various metrics. Simulation results and corresponding discussions are provided in Section VI. Finally, Section VII concludes the paper.

\textit{Notations:} Throughout this paper, boldface lowercase and uppercase letters denote vectors and matrices, respectively. For a complex-valued matrix $\mathbf{A}\in\mathbb{C} ^{M\times N}$, $\mathbf{A}^T$, $\mathbf{A}^H$ and $\mathbf{A}_{[m,n]}$ represent its transpose, Hermitian transpose, and the element in the $m$-th row and $n$-th column, respectively. For a \textit{random variable} (RV) $X$, $\mathbb{E}\{X\}$, $\mathbb{E} \{X^2\}$, $\text{Var}\{X\} = \mathbb{E}\{X^2\} - \mathbb{E}^2 \{X\}$ and $\mathbb{P} \{X >T \}$ denote its expectation, second-order moment, variance and the probability that $X$ exceeds a threshold $T$, respectively. The notations $X\sim \mathcal C\mathcal N(\mu, \sigma^2)$, $X\sim \Gamma(\alpha , \beta)$, $X\sim \text{Exp}(\lambda)$ and $X\sim B(a,b)$ indicate that $X$ follows a \textit{circularly symmetric complex Gaussian} (CSCG), Gamma, exponential or binomial distributions, where $\mu$ and $\sigma^2$ are the mean and variance for the CSCG distribution, $\alpha $ and $\beta $ are the shape and scale parameters of the Gamma distribution, $\lambda$ is the intensity parameter of the exponential distribution, and $a$ and $b$ are the parameters of the binomial distribution. 

\section{Network Framework and Signal Model}
\begin{figure}[t]
  \centering
  \includegraphics[scale = 0.5]{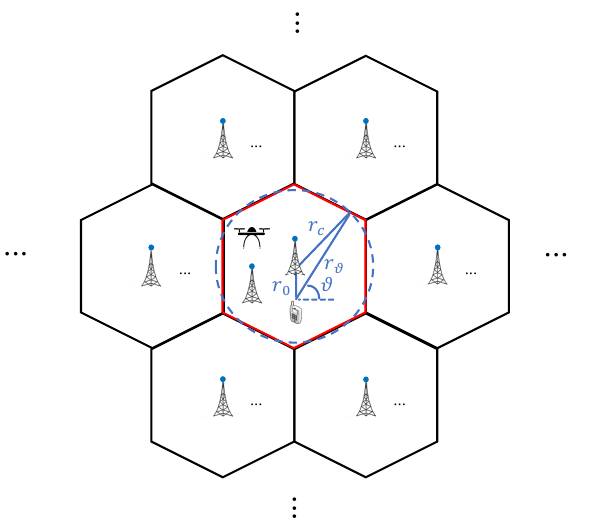}
  \caption{Geometric topology for the CoMP ISAC network.}
  \label{FigCoopSys}
\end{figure}

This paper investigates a heterogeneous air-ground ISAC network with a CoMP architecture, in which BSs, CUs, and STs are spatially distributed according to independent stationary point processes. The deployment of BSs adopts a hybrid model: master BSs are placed in a conventional hexagonal lattice layout, while slave BSs are randomly distributed following a \textit{two-dimensional} (2-D) HPPP $\Phi_B$ with intensity $\lambda_B \ \mathrm{BSs/Km}^2$. The master BSs, equipped with abundant resource and global network awareness, serve as central units that coordinate centralized C\&S operations for CUs and STs. Meanwhile, the slave BSs act as distributed radio access nodes, enhancing performance through coordinated C\&S functionalities that leverage spatial diversity across the network.

As depicted in Fig.~\ref{FigCoopSys}, the network is partitioned into multiple geographically disjoint cooperative clusters. Each cluster consists of a central master BS, located at the lattice site, along with multiple slave BSs uniformly distributed within the corresponding hexagonal region. To ensure analytical tractability, a total power constraint is imposed per cluster, with each BS transmitting at an average power of $P_t$. The model assumes ideal front-haul links between slave and master BSs, full CSI availability within the cluster at the master BS, and perfect synchronization across the network, while disregarding practical limitations related to CSI acquisition and data exchange. These assumptions allow the application of localized \textit{cell-free} (CF) MIMO techniques for ISAC, thereby enabling the derivation of theoretically optimal performance bounds. Moreover, for the computation of both desired signal and aggregated interference power, the typical hexagonal cluster is approximated by a circle of radius $r_c$ preserving equal area, as illustrated in Fig.~\ref{FigCoopSys}.

\begin{figure}[t]
  \centering
  \includegraphics[scale = 0.5]{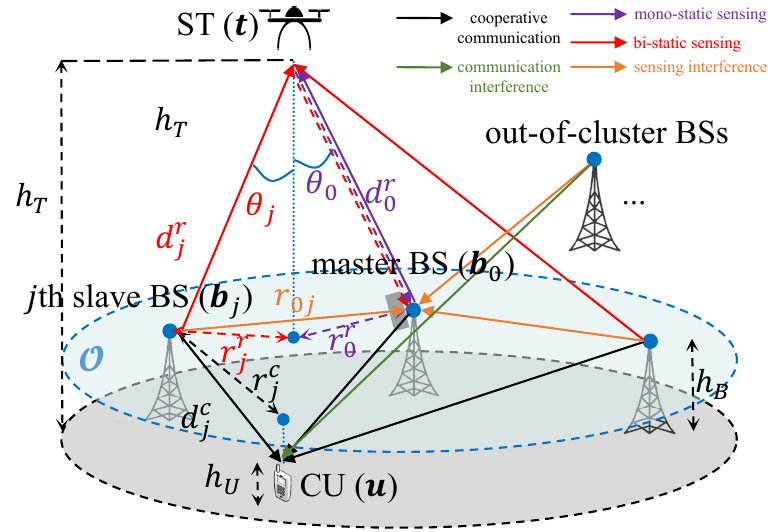}
  \caption{Geometry model for the air-ground ISAC network.}
  \label{FigSys}
\end{figure}

\subsection{ISAC Network Model}
Consider a downlink ISAC network organized into cooperative clusters. The $l$-th cluster consists of a master BS, denoted as $\bm b_0$, and $N_l$ slave BSs modeled by a finite PPP $\Phi_S = \{\bm b_j\}_{j=1:N_l}$. Each BS is equipped with a vertically oriented \textit{uniform linear array} (ULA) of $N_t$ transmit antennas, while the master BS is additionally equipped with a dedicated ULA of $N_r$ receive antennas\footnote{With the ULA placed vertically, the orientation of the BS does not affect the position of each antenna.}. 

Within each transmission interval of interesting, the cluster serves $\eta_cN_tN_l$ single-antenna and uniformly distributed CUs, and senses $\eta_r N_t N_l$ point-like aerial STs, where $\eta_c \in \left[0, 1\right] $ and $\eta_r = 1 - \eta_c$ represent the C\&S loading factors, respectively. Following the SG framework, a typical CU and a typical ST are analyzed to evaluate the network's average performance \cite{andrews2011tractable}. 

The master BS is assumed to possess perfect knowledge of the relative parameters between itself and the ST, obtained through preliminary mono-static sensing in dedicated subframes. In this case, one can focus on the bi-static sensing and further cooperative hybrid mono/bi-static sensing to obtain the performance upper bound of the system. Specifically, under the CoMP transmission scheme, all slave BSs within the cluster synchronously transmit identical ISAC waveforms to enable joint communication and bi-static sensing. The master BS acts as a centralized receiver, processing the reflected signals from the cooperative bi-static sensing links. 

Taking the horizontal plane $\mathcal{O}$ where BSs are located as the reference, the geometric relationships are illustrated in Fig. \ref{FigSys}. The BSs, CUs, and STs are positioned at heights $h_B \ \mathrm{m}$, $h_U \ \mathrm{m}$ and $h_T \ \mathrm{m}$, respectively, with $h_T>h_B>h_U $ assumed without loss of generality. Let $\Delta h_c = h_B - h_U$ denotes the vertical distance between BSs and CUs, and $\Delta h_r = h_T - h_B$ that between STs and BSs. The horizontal distances from the $j$-th BS to the typical CU and ST are defined as $\left\{r_j^{c} = \left\lVert \bm u - \bm b_j\right\rVert\right\} $ and $\left\{r_j^{r} = \left\lVert \bm t - \bm b_j\right\rVert\right\} $, respectively. The corresponding \textit{three-dimensional} (3-D) link distances are given by $d_j^{c} = \sqrt{\left(r_j^{c}\right) ^2 + \Delta h_c^2 }$ and $d_j^{r} = \sqrt{\left(r_j^{r}\right) ^2 + \Delta h_r^2 }$, respectively. The spatial angle between BS $j$ and the typical ST is denoted as $\theta_j$, and the horizontal inter-BS distance between BS 0 and BS $j$ is $r_{0j} = \left\lVert \bm b_0 - \bm b_j\right\rVert $. For notational simplicity in subsequent analysis, $r_j$ and $d_j$ are used to represent the 2-D horizontal and 3-D link distances, respectively, retaining only the subscript $j$. This simplification is unambiguous as the terms are clearly distinguished in the contexts of C\&S, respectively.

\subsection{Signal Model}
The considered ISAC network employs OFDM signaling due to its excellent performance in both C\&S. The system operates at a carrier frequency $f_c$ with a total bandwidth $B$, which is divided into $N$ subcarriers with spacing $\Delta f$. Each OFDM frame consists of $M$ symbols, and the symbol duration is $T = T_g + T_s$, where $T_g$ and $T_s$ represent the guard interval and data symbols duration, respectively, leading to a subcarrier spacing $\Delta f = 1/T_s$. The channel is assumed to remain stationary over the $M\times N$ time-frequency resource blocks in an OFDM frame.

\subsubsection{Communication Model}
To ensure fair interference distribution among CUs, each CU is randomly assigned $N/(\eta_cN_tN_l)$ subcarriers for data transmission\footnote{As resource allocations are beyond the scope of our work, a simplified FDM scheme is adopted where resources are allocated on a per-subcarrier basis. This serves as a tractable baseline, and more advanced schemes considering both subcarriers and OFDM symbols may lead to better performance.}. Following established methodologies \cite{olson2023coverage, lin2015performance, ali2019stochastic}, the communication performance is evaluated on an arbitrary RE basis. 

Let $\bm g_{ij} = \sqrt{\mathcal{L} _{ij}^c}\bm h_{ij} \in \mathbb{C}^{N_t\times 1 }$ denote the channel vector from the $j$-th BS in the $i$-th cluster to the typical CU, where $\mathcal{L} _{ij}^c = \left(\frac{\lambda_c}{4\pi}\right)^2 d_{ij}^{- \alpha_c } $ is the path loss with exponent $\alpha_c$ over distance $d_{ij}$, $\lambda_c$ is the carrier wavelength, and $\bm h_{ij} \sim \mathcal{C} \mathcal{N} (\bm 0, \bm I_{N_t})$. The collective communication channel from all $N_l$ cooperative BSs in the serving cluster $l$ is given by $\bm g_l = \left[\bm g^T_{l1}, \cdots, \bm g^T_{lj}, \cdots, \bm g^T_{lN_l} \right]^T \in \mathbb{C} ^{N_tN_l \times 1}$. Similarly, the collective communication interference channel from all $N_i$ BSs in an interfering cluster $i$ is $\bm f_i = \left[\bm g^T_{i1}, \cdots, \bm g^T_{ij}, \cdots, \bm g^T_{iN_i} \right]^T \in \mathbb{C} ^{N_iN_t \times 1}$. 

Assuming equal power allocation across all REs for the $\eta_cN_tN_l$ CUs, the received signal at the typical CU can be expressed as  
\begin{equation}
  \begin{aligned}
    \label{comm signal}
  y_c &= \underbrace{ \sqrt{P_t }\bm g^H_l \bm w_l s_l}_{\text{intended signal}} + \underbrace{\sum\nolimits_{i \neq l } \sqrt{P_t} \bm f^H_i \bm w_i s_i }_{\text{inter-cluster interference} } + z_c ,
\end{aligned}
\end{equation}
where $\bm w_i = \left[\bm w^T_{i1}, \cdots, \bm w^T_{ij}, \cdots, \bm w^T_{iN_l} \right]^T \in \mathbb{C} ^{N_iN_t \times 1}$ represents the normalized BF vector in cluster $i$, $s_i$ is the transmitted symbol from cluster $i$, and $z_c\sim \mathcal{C}\mathcal{N}(0, \sigma_c^2 )$ is additive white Gaussian noise. 

\subsubsection{Sensing Model}
The sensing model operates on a fundamentally different paradigm from communication. Unlike communication which focuses on data transmission within specific REs, radar sensing requires processing echo signals across the entire time-frequency resources within a \textit{coherent processing interval} (CPI) to extract target parameters that are inherently coupled across temporal, spectral, and spatial domains, with performance positively correlated with available resources. Our framework adopts a cooperative hybrid mono/bi-static sensing scheme under \textit{line-of-sight} (LoS) conditions, wherein the master BS acts as a centralized receiver processing target echos from both itself and cooperating slave BSs. Key sensing tasks include target detection and parameter estimation, with parameters of interest including the relative distance $d_j = c\tau_j/2$ for each bi-static sensing link $j$ and the real ST velocity $v_{\text{real}}$, where $c$ is the speed of the light and $\tau_j$ is the time delay. 

The estimation of distance and speed in OFDM-radar systems can be formulated as a spectral estimation problem using periodogram algorithm \cite{sturm2011waveform, Braun14}. This allows separate parameter estimation for each ST once the number of STs is determined. Under the assumption that different STs are distinguishable in the pseudo-spectrum and only subject to background interference and noise, one can focus on a typical ST for performance analysis. The radar observation for the typical ST on the $(n,m)$-th RE is given by \eqref{raw radar observation}\footnote{Intra-cluster interference is assumed to be canceled by cooperative BF. Also, unlike coherent merging in a communications context that operates on clusters, the cluster indices are omitted here since non-coherent sensing processing is being considered.},
\begin{figure*}[!t]
\begin{align}
    \label{raw radar observation}
  \left[\bm Y_r\right]_{n,m} &= \underbrace{\sum_{j\in\Phi_S} \sqrt{P_t \mathcal{L}_{0j}^r} \bm a^H(\theta_0) \underbrace{\bm a_r(\theta_0)\bm a_t^T(\theta_j)}_{\triangleq \bm G_{0j}(\theta_0, \theta_j) } \bm w_{lj} \left[\bm S_j \right]_{n,m} e^{\jmath 2\pi m T f_{D_{0j}} } e^{-\jmath 2\pi n \Delta f\tau_{0j} }}_{\text{distributed bi-static sensing}} \nonumber \\
  & + \underbrace{ \sum_{j \in \Phi_I \triangleq \Phi_B | \Phi_S } \sqrt{P_t \mathcal{L}_{0j}^c} \underbrace{\bm a^H(\theta_0) \bm H _{0j}}_{ \triangleq \bm \nu_{0j}^H(\theta_0)} \bm w_{ij} \left[\bm S_j\right]_{n,m} e^{\jmath \beta_{0j} } }_{\text{inter-cluster interference} } + \left[\bm Z_r\right]_{n,m} ,
\end{align}
\vspace{-0.3cm}
\end{figure*}
where $\mathcal{L}_{0j}^r = \frac{\lambda_c^2\xi }{(4\pi)^3} (d_0  d_j)^{-\alpha _r} $ is the bi-static sensing path loss, with $\xi$ denoting the \textit{radar cross-section} (RCS) and $\alpha _r$ the path loss exponent, $\bm a^H(\theta_0) \in \mathbb{C} ^{1\times N_r}$ is the receive BF vector, and $\bm G_{0j}(\theta_0, \theta_j) \triangleq \bm a_r(\theta_0)\bm a_t^T(\theta_j)$ combines the steering vectors $\bm a_r(\theta_0) = [1, \cdots, e^{\jmath \pi (N_r-1)\cos \theta_0}]^T \in \mathbb{C} ^{N_r\times 1}$ as well as $\bm a_t(\theta_j) = [1, \cdots, e^{\jmath \pi (N_t-1)\cos \theta_j}]^T \in \mathbb{C} ^{N_t \times 1}$, and $f_{D_{0j}}$ and $\tau_{0j}$ corresponding to the estimated parameters for the $j$-th bi-static sensing link. The sensing interference components comprise three elements: (1) the path loss between interfering BS $j$ and the master BS, given by $\mathcal{L}_{0j}^c = \left(\frac{\lambda_c}{4\pi}\right)^2 r_{0j}^{- \alpha_c } $; (2) the interference channel matrix $\bm H _{0j} \in \mathbb{C} ^{N_r \times N_t}$ with i.i.d. $\mathcal{C}\mathcal{N}(0, 1)$ entries; and (3) the effective sensing interference channel $\bm \nu_{0j} ^H(\theta_0) = \bm a^H(\theta_0) \bm H _{0j} \in \mathbb{C} ^{1 \times N_t}$ under receive BF vector $\bm a^H(\theta_0)$. The system noise is modeled as additive white Gaussian noise denoted by $\left[\bm Z_r\right]_{n,m} \sim \mathcal{C}\mathcal{N}( 0, \sigma_r^2 )$. The BF vectors $\bm w_{lj}$ and $\bm w_{ij}$ depend solely on quasi-static channel conditions, permitting omission of $[n,m]$ subscripts in statistical analysis.
  
To decouple the impact of the transmitted symbol $\left[\bm S_0\right]_{n,m}$ on sensing, the symbol-by-symbol phase rotation is applied, yielding the compensated sensing signal in \eqref{sensing signal} at the top of this page,
\begin{figure*}[!t]
\begin{align}
  \label{sensing signal}
  \left[\widetilde{\bm Y} _r\right]_{n,m} &= \underbrace{\sum_{j \in \Phi_S} \sqrt{P_t \mathcal{L}_{0j}^r} \bm a^H(\theta_0) \bm G_{0j}(\theta_0, \theta_j) \bm w_{lj} e^{\jmath 2\pi m T {f_D}_{0j} } e^{-\jmath 2\pi n \Delta f \tau_{0j} }}_{\text{distributed bi-static sensing}} \nonumber \\
  & + \underbrace{ \sum_{j \in \Phi_I } \sqrt{P_t \mathcal{L}_{0j}^c } \bm \nu ^H_{0j}(\theta_0) \bm w_{ij} e^{\jmath \widetilde{\beta} _{0j}} }_{\text{inter-cluster interference} } + \left[\widetilde{\bm Z} _r\right]_{n,m} ,
\end{align}
\hrulefill
\vspace{-0.4cm}
\end{figure*}
where $\widetilde{\beta} _{0j}$ is the residual interference phase and $\left[\widetilde{\bm Z} _r\right]_{n,m}$ is the equivalent noise after compensation. The constant-amplitude modulation ($\left\lvert s_j \right\rvert = 1$) ensures the statistical properties of both reference signal and noise remain invariant under phase rotation.

\subsection{Cooperative BF Design}
To mitigate the significant power imbalance between bi-static sensing echoes, which undergo dual-path attenuation, and the strong interference directly from co-cluster slave BSs, a cooperative BF scheme is implemented to suppress intra-cluster interference. The transmitter employs a linear \textit{zero-forcing} (ZF) BF strategy \cite{yoo2006optimality} to achieve spatial diversity while preserving analytical tractability. 

On the other hand, at the receiver, leveraging prior knowledge of the direction $\theta _{0k}$ between the $k$-th ST and the master BS obtained from previous sensing measurements, the \textit{maximum ratio combining} (MRC) receive BF vector is designed as
\begin{equation}
  \bm a^H(\theta_{0k}) = \frac{1}{\sqrt{N_r}}[1, \cdots, e^{-\jmath \pi (N_r-1)\cos \theta_{0k}} ]\in \mathbb{C} ^{1\times N_r} ,
  \end{equation}
which maximizes the received power of the sensing echo. Notably, since linear combinations of complex Gaussian RVs remain Gaussian, the effective sensing interference channel vector $\bm \nu _{0j}^H(\theta_{0k})$ retains Rayleigh distribution. 

Leveraging this property, cooperative ZF BF is implemented at the $l$-th cluster to nullify the intra-cluster sensing interference from the $N_l$ slave BSs to the master BS. This is achieved by projecting the BF vector $\bm w_l$ onto the null space of the collective effective sensing interference channels, which is constructed as
\begin{equation}
  \begin{aligned}
    \bm \nu_{lk} = &\left[\bm \nu_{01}^H(\theta_{0k}), \bm \nu_{02} ^H(\theta_{0k}), \cdots, \bm \nu_{0N_l} ^H(\theta_{0k}) \right]^T \in \mathbb{C} ^{N_tN_l\times 1}, \\
    & \forall k \in \left\{1, \cdots, \eta_r N_tN_l\right\}  .
  \end{aligned}
\end{equation}
Mathematically, the normalized cooperative ZF BF vector for cluster $l$ can be given by
\begin{equation}
 \bm w_l = \frac{\left(\bm I_{N_tN_l} - \bm G_l \bm G^\dagger _l \right)\widehat{\bm g}_l }{ \left\lVert \left(\bm I_{N_tN_l} - \bm G_l \bm G^\dagger _l \right)\widehat{\bm g}_l \right\rVert } \in \mathbb{C} ^{N_tN_l\times 1}  ,
\end{equation} 
where the composite channel matrix $\bm G_l = \left[\widehat{\bm \nu}_{l1} , \cdots, \widehat{\bm \nu}_{l\eta_rN_tN_l} \right]^T $ stacks the normalized effective sensing interference channels $\widehat{\bm \nu}_{lk} = \frac{\bm \nu_{lk}}{\left\lVert \bm \nu_{lk}\right\rVert }$ for all STs, and $\widehat{\bm g}_l = \frac{\bm g_l}{\left\lVert \bm g_l\right\rVert }$. Due to the orthogonality inherent in ZF BF, the dimension of the BF subspace, i.e., the spatial diversity order provided to the typical CU is $\zeta _l = N_tN_l - \eta _r N_tN_l = \eta _c N_tN_l $.

\section{Distribution Analysis}
As indicated by the signal model, the received signal quality is directly influenced by the power of both the desired channel and the interference channel after projection onto the BF subspace. These powers depend on the large-scale path loss and the per-BS BF gain, the latter being defined in terms of the small-scale fading channel. Accordingly, this section first derives the statistical distributions of the per-BS BF gains. Subsequently, the necessary distributions of the relevant link distances are characterized. Finally, as a key intermediate step for subsequent performance analysis, the Laplace transforms of the aggregated interference and desired signal powers are derived, which serve as fundamental tools for performance analysis.

\subsection{Signal Power Distribution}
To achieve analytical tractability, a Gamma second-order moment matching is employed to interpret the intended signal and aggregated interference signal powers as projections of isotropic channels, yielding the following approximated distributions for C\&S, as summarized in Corollary 1 proved in Appendix A and Corollary 2 proved in Appendix C in the supplemental materials, respectively.

\textit{\textbf{Corollary 1 (Communication Signal and Interference Power Distributions):}} In the considered air-ground ISAC CoMP network, the effective communication signal power and the aggregated communication interference power at a typical CU can be approximated via Gamma distributions, interpreted as projections from isotropic channels. Specifically, 
\begin{subequations}
  \begin{align}
    &\left\lvert \bm g^H _l\bm w_l\right\rvert^2 \overset{d}{= } \sum_{j =1}^{N_l} \mathcal{L} _{lj}^c g_{lj}^{cs} = \sum_{j \in \Phi_S} \mathcal{L} _j^c g_{j}^{cs}  , \\
    &\sum_{i \neq l } \left\lvert \bm f^H_i \bm w_i \right\rvert^2 \overset{d}{= } \sum_{i \neq l } \sum_{j = 1}^{N_i} \mathcal{L} _{ij}^c g_{ij}^{ci}   ,
  \end{align}
\end{subequations}  
where 
\begin{subequations}
  \begin{align}
    g_j^{cs} &\sim \Gamma \left(\eta _c N_t \triangleq m_c, 1 \right) , \\
    g_{ij}^{ci} &\sim \Gamma \left(\frac{1}{N_i}, 1 \right)
  \end{align}
\end{subequations}  
represent the BF gains for the effective communication channel and the communication interference channel, respectively.

\textit{\textbf{Remark 1:}} The BF gain for the communication interference channel, $g_{ij}^{ci}$, depends on the specific number of slave BSs within each cluster. To maintain analytical tractability, a common approach adopted in \cite{hosseini2016stochastic} is to replace $N_i$ with the average number of slave BSs per cluster. However, this averaging method often introduces discrepancies between analytical and simulation results, as it ignores the inherent randomness of the PPP. To address this, a fact is leveraged that conditioned on the number of slave BSs $N_i$, a finite PPP is equivalent to a \textit{binomial point process} (BPP). Therefore, the statistical performance metrics for the finite PPP model can be derived by first obtaining conditional metrics under the BPP assumption (with fixed $N_l$ BSs), and then averaging over the Poisson-distributed $N_l$. Based on this insight, a BPP-based modeling approach is adopted, where the number of slave BSs per cluster is fixed ($N_i = N_l$ for $\forall i$) to facilitate tractability in subsequent analysis. Under this model, the set of co-cluster slave BSs constitutes a BPP with $N_l$ points, while the remaining interfering BSs asymptotically converge to a PPP with equivalent density, due to the Poisson limit theorem for BPPs. Consequently, the aggregated interference satisfies $\sum_{i \neq l } \sum_{j = 1}^{N_i} \mathcal{L} _{ij}^c g_{ij}^{ci} = \sum_{i \neq l } \sum_{j = 1}^{N_l} \mathcal{L} _{ij}^c g_{ij}^{ci} \overset{d}{\approx } \sum_{j \in \Phi_I} \mathcal{L} _j^c g_{j}^{ci}$, where $g_j^{ci} \sim \Gamma \left(\frac{1}{N_l }, 1 \right) $.

In this work, the interference-limited networks are analyzed where noise is negligible due to dominant interference in dense cellular deployments. Accordingly, the \textit{signal-to-interference ratio} (SIR) is adopted as the performance metric \cite{lee2014spectral, park2016optimal}. Combining the communication signal model with the preceding analysis, the received SIR at the typical CU can be expressed as
\begin{equation}
  \begin{aligned}
    \label{communication SIR}
    \gamma _c = \frac{ \left\lvert \bm g^H _l\bm w_l\right\rvert^2}{ \sum_{i \neq l } \left\lvert \bm f^H_i \bm w_i \right\rvert^2  } \overset{d}{\approx } \frac{\sum_{j \in \Phi_S} \mathcal{L} _j^c g_{j}^{cs}}{\sum_{j \in \Phi_I} \mathcal{L} _j^c g_{j}^{ci}} = \frac{\underbrace{\sum\nolimits_{j \in \Phi_S} d _j^{- \alpha _c} g_{j}^{cs}}_{\triangleq S_{ac}} }{\overbrace{\sum\nolimits_{j \in \Phi_I} d _j^{- \alpha _c} g_{j}^{ci}}^{\triangleq I_{ac}} } .
\end{aligned}
\end{equation} 

In contrast to communication signals, which benefit from coherent merging, radar sensing echoes from each bi-static link exhibit independent phases due to distinct observation geometries, necessitating a per-link analysis. To characterize the sensing BF gains, two foundational lemmas are introduced as proved in Appendix B in the supplemental materials.

\textit{\textbf{Lemma 1 (Gamma 2nd-Order Moment Matching):}} Let $X$ be a RV with expectation $\rho = \mathbb{E} \left\{X \right\}$, second moment $\rho^{(2)} = \mathbb{E} \left\{X^2 \right\}$, and variance $\varrho = \rho^{(2)} - \rho^2$. Then $X$ can be approximated by a Gamma distribution $\Gamma(\alpha , \beta )$ with shape and scale parameters:
\begin{equation}
\alpha = \rho^2/\varrho \ \text{and} \ \beta = \varrho/\rho .
\end{equation}

\textit{\textbf{Lemma 2 (Expectations over Isotropic Random Vectors):}} Let $\bm w_l$ be a normalized $N_tN_l \times 1$ isotropic random vector and $\bm a_t^T(\theta_j)$ is a constant steering vector. Then, one can get 
\begin{subequations}
  \begin{align}
  &\rho _{| \bm a_t^T(\theta_j)} := \mathbb{E}_{\bm w_{lj}} \left\{\left\lvert \bm a_t^T(\theta_j) \bm w_{lj} \right\rvert^2 \right\} 
  = \frac{\bm a_t^H(\theta_j) \bm a_t(\theta_j)}{N_tN_l} , \\
&\varrho_ {| \bm a_t^T(\theta_j)} := \text{Var}_{\bm w_{lj}} \left\{\left\lvert \bm a_t^T(\theta_j) \bm w_{lj} \right\rvert^2 \right\} = \frac{N_t - 1}{N_t + 1} \rho^2 _{| \bm a_t^T(\theta_j)} .
\end{align}
\end{subequations}

\textit{\textbf{Corollary 2 (Sensing Signal BF Gain Distribution):}} In the considered air-ground ISAC CoMP network, the BF gain for the effective sensing channel admits a Gamma approximation: 
\begin{equation}
  \begin{aligned}
    \left\lvert \bm a_t^T(\theta_j) \bm w_{lj} \right\rvert^2 \triangleq g_j^{rs}  ,
  \end{aligned}
\end{equation}  
where 
\begin{equation}
  \begin{aligned}
    g_j^{rs} \sim \Gamma\left(\frac{N_t + 1}{N_t - 1}, \frac{N_t - 1}{(N_t + 1)N_l}\right)  .
  \end{aligned}
\end{equation}  

\textit{\textbf{Remark 2:}} To enhance tractability, it is approximated that $g_j^{rs} \sim \Gamma(1, 1/N_l) \overset{d}{= } \text{Exp}(N_l)$ when the number of transmit antennas $N_t$ is sufficiently large. Furthermore, since the effective sensing interference channel vector $\bm \nu_{0j}^H(\theta_0)$ is Rayleigh distributed, and its linear combination through the BF vector $\bm w_{lj}$ results in a Gaussian variable whose power is scaled by $1/N_l$, the BF gain for the sensing interference channel also follows an exponential distribution: $\left\lvert \bm \nu _{0j}^H(\theta_0) \bm w_{ij}\right\rvert^2 \triangleq g_j^{ri} \sim \text{Exp}(N_l)$. 

From the radar sensing echo in \eqref{sensing signal}, the sensing SIR for the $j$-th bi-static sensing link is\footnote{Note that for both C\&S SIRs in \eqref{communication SIR} and \eqref{sensing SIR}, the transmit power $P_t$ is canceled out. This indicates that the results provide fundamental performance insights under the considered equal-power model, while further power allocation may enhance performance and is reserved for future work.}
\begin{equation}
\begin{aligned}
  \label{sensing SIR}
  \gamma _{0j}^r = \frac{ \mathcal{L}_{0j}^r N_r \left\lvert \bm a_t^T(\theta_j) \bm w_{lj}\right\rvert^2 }{\sum_{j \in \Phi_I } \mathcal{L}_{0j}^c \left\lvert \bm \nu _{0j}^H(\theta_0) \bm w_{ij}\right\rvert^2 } \overset{d}{= } \frac{\xi (4\pi)^{-1} (d_0 d_j)^{-\alpha _r} N_r g_j^{rs}}{\underbrace{\sum\nolimits_{j \in \Phi_I } r_{0j}^{-\alpha _c} g_j^{ri}}_{\triangleq I_{ar}} } ,
\end{aligned}
\end{equation} 
where $N_r$ comes from the MRC gain, which depends directly on the number of receive antennas.

\begin{figure}[t]
  \centering
  \subfigure[Case 1]{
  \includegraphics[scale=0.5]{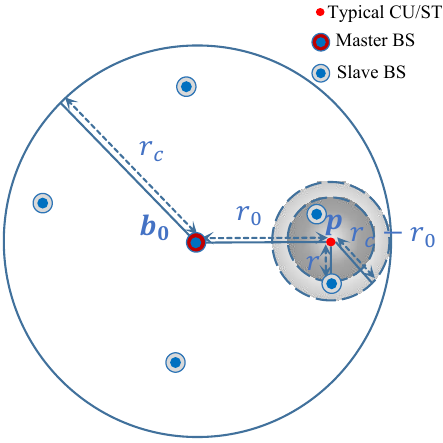}}
  \subfigure[Case 2]{
  \includegraphics[scale=0.5]{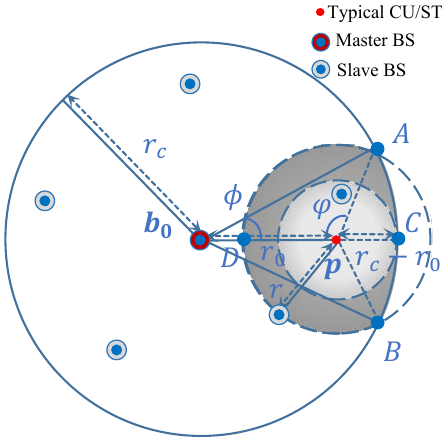}}
  \caption{Illustration of sensing serving distance under different cases, wherein (a) Case 1: $0\leq r < r_c - r_0$, and in (b) Case 2: $r_c - r_0 \leq r \leq r_c + r_0$.}
  \label{Distance Model}
\end{figure}

\subsection{Distance Distribution}
As illustrated in Fig. \ref{Distance Model}, the master BS is located at $\bm b_0 = (0, 0)$, and a set of $N_l$ cooperative slave BSs, denoted by $\Phi_S = \{\bm b_j\}_{j=1:N_l}$, are independently and uniformly distributed within a disk $\mathcal{D}(\bm b_0, r_c)$ centered at the master BS with radius $r_c$. The typical CU/ST is located at $\bm p = (r_0, 0)$. The \textit{probability density function} (PDF) of the distance $r_0$ is given by
\begin{equation}
  \label{pdf r0}
  f_{R_0}(r_0) = \frac{2r_0}{r_c^2} \mathbf{1}\left(0\leq r_0 \leq r_c \right) .
\end{equation}

Note that the distances from cooperative slave BSs to the typical CU/ST, denoted by $\left\{R_j = \left\lVert \bm b_j - \bm p\right\rVert\right\}_{j = 1:N_l } $, are correlated due to common factor $\bm p$ (i.e., $r_0$). However, conditioned on $r_0$, the set $\left\{R_j\right\}_{j = 1:N_l }$ becomes conditionally i.i.d. The conditional sampling \textit{cumulative distribution function} (CDF) and PDF of each element in $\left\{R_j \right\}_{j = 1:N_l } $ are provided in the following lemma as approved in Appendix D in the supplemental materials. 

\textit{\textbf{Lemma 3 (Conditional CDF and PDF of Sampling Distance $R$):}} Conditioned on the distance $r_0$, the sampling CDF of the unordered horizontal distance $R$ from an arbitrary slave BS to the typical CU/ST is
\begin{equation}
  \begin{aligned}
    &F_{R}(r|r_0) = \frac{r^2}{r_c^2} \mathbf{1}\left(0\leq r< r_c - r_0\right) +\\
    &\frac{\varphi r^2 + \phi r_c^2- r_0r_c\sin \phi}{\pi r_c^2}  \mathbf{1}\left(r_c - r_0\leq r\leq r_c + r_0\right),
  \end{aligned}
\end{equation}
and the corresponding conditional PDF is
\begin{equation}
  \begin{aligned}
    \label{unordered distance pdf}
    f_{R}(r|r_0) &= \frac{2r}{r_c^2}\mathbf{1}\left(0\leq r< r_c - r_0\right) + \frac{2\varphi r}{\pi r_c^2} \\
    &\times \mathbf{1}\left(r_c - r_0\leq r\leq r_c + r_0\right) ,
  \end{aligned}
\end{equation}
where $\phi = \arccos \left(\frac{r_c^2 + r_0^2 - r^2}{2r_cr_0}\right) $, $\varphi = \arccos \left(\frac{r^2 + r_0^2 - r_c^2}{2r_0r}\right) $, and $\mathbf{1}(\cdot)$ denotes the indicator function.

The ordered set $\left\{R_{j:N_l}\right\}_{j = 1:N_l }$ is defined by sorting the elements of the unordered set $\left\{R_j\right\}_{j = 1:N_l }$ in ascending order, such that $R_{j:N_l}$ denotes the distance from the $j$-th nearest slave BS to the typical CU/ST. The conditional PDF of this ordered distance is given in the following corollary as proved in Appendix E in the supplemental materials.

\textit{\textbf{Corollary 3 (Conditional CDF and PDF of Ordered Distance $R_j$):}} Conditioned on the distance $r_0$, the conditional PDF of ordered horizontal distance from the $j$-th nearest BS to the typical CU/ST $R_j$ is given as
\begin{equation}
  \begin{aligned}
    f_{R_j}(r_j|r_0) &= \frac{(1 - F_{R}(r_j|r_0))^{N_l - j}(F_{R}(r_j|r_0))^{j-1} }{\mathcal{B} (N_l - j + 1, j)} f_{R}(r_j|r_0) ,
  \end{aligned}
\end{equation}
where $\mathcal{B}(a,b)$ denotes the beta function, expressible in terms of gamma functions as $\mathcal{B}(a,b) = \Gamma (a)\Gamma(b)/\Gamma(a+b)$.

\subsection{Laplace Transforms}
In this subsection, the Laplace transform of different signal power are introduced in following propositions as proved in Appendix F - I in the supplemental materials.

\textit{\textbf{Proposition 1 (Laplace Transform of Intended Communication Signal Power $S_{ac}$ under PPP Model):}} Under the PPP model, the Laplace transform of the aggregated intended communication signal power, conditioned on the distance $R_0$ from the master BS to the typical CU, is given by
\begin{equation}
  \begin{aligned}
    \mathcal {LT}_{S_{ac} |R_0 }(t) = e^{\lambda_B \int_{0}^{2\pi} \mathcal{I}_1^{cs}(\vartheta) \,d\vartheta } ,
  \end{aligned}
\end{equation}
where  
\begin{equation}
  \begin{aligned}
    &\mathcal{I}_1^{cs}(\vartheta) = \frac{ r_\vartheta^2 + \Delta h_c^2 }{2}\\
    &\cdot \left[_2F_1 \left(m_c, - \frac{2}{\alpha_c}; 1 - \frac{2}{\alpha_c}; - t \left(r_\vartheta^2 + \Delta h_c^2\right)^{- \frac{\alpha _c}{2}}\right)  - 1 \right] \\
    & - \frac{\Delta h_c^2}{2} \left[_2F_1 \left(m_c, - \frac{2}{\alpha_c}; 1 - \frac{2}{\alpha_c}; - t \Delta h_c^{-\alpha_c}\right)  - 1 \right] ,
\end{aligned}
\end{equation}
with 
\begin{equation}
  \begin{aligned}
    r_\vartheta = \sqrt{r_c^2 - r_0^2\cos^2 \vartheta } + r_0\sin \vartheta
  \end{aligned}
\end{equation} 
representing the distance from the CU to the cluster boundary along the radial direction at angle $\vartheta$ and $_2F_1(a , b; c; z ) = F(a , b; c; z )$ denoting the Gauss hypergeometric function \cite{gradshteyn2014table}.
 
\textit{\textbf{Proposition 2 (Laplace Transform of Intended Communication Signal Power $S_{ac}$ under BPP Model):}} Under the BPP model, the Laplace transform of the aggregated intended communication signal power, conditioned on $R_0$, is given by
\begin{equation}
  \begin{aligned}
    \mathcal {LT}_{S_{ac} |R_0 }(t) = \left[\Psi_1 (t) - \Psi_2 (t) + \mathcal{I}_2^{cs}(t)\right]^{N_l}  ,
  \end{aligned}
\end{equation}
where  
\begin{subequations}
  \begin{align}
&\Psi_1 (t) = \frac{(r_c - r_0)^2 + \Delta h_c^2}{r_c^2} \\
&\cdot {_2F_1} \left(m_c, - \frac{2}{\alpha_c}; 1 - \frac{2}{\alpha_c}; - t \left[(r_c - r_0)^2 + \Delta h_c^2\right]^{- \frac{\alpha_c}{2}} \right) \nonumber, \\
&\Psi_2 (t) = \frac{\Delta h_c^2}{r_c^2} \ {_2F_1} \left(m_c, - \frac{2}{\alpha_c}; 1 - \frac{2}{\alpha_c}; - t \Delta h_c^{- \alpha_c}\right)  ,\\
&\mathcal{I}_2^{cs}(t) = \int_{r_c - r_0}^{r_c + r_0} \left[1 + t \left(r^2 + \Delta h_c^2 \right) ^{-\frac{\alpha _c}{2}}\right] ^{-m_c} \frac{2r}{\pi r_c^2} \nonumber \\
&\cdot \arccos \left(\frac{r^2 + r_0^2 - r_c^2}{2r_0r}\right) \,dr .
  \end{align}
\end{subequations}

\textit{\textbf{Proposition 3 (Laplace Transform of Aggregated Communication Interference Power $I_{ac}$):}} The Laplace transform of the aggregated communication interference power, conditioned on $R_0$, is given by
\begin{equation}
  \begin{aligned}
    \mathcal {LT}_{I_{ac} |R_0 }(t) = e^{\lambda_B \int_{0}^{2\pi} \mathcal{I}_1^{ci}(\vartheta) \,d\vartheta } ,
  \end{aligned}
\end{equation}
where
\begin{equation}
  \begin{aligned}
\mathcal{I}_1^{ci}(\vartheta) &= \frac{r_\vartheta^2 + \Delta h_c^2}{ 2} \left[ 1 - \  _2F_1\left(\frac{1}{N_l}, - \frac{2}{\alpha _c}; 1 - \frac{2}{\alpha _c}; \right.\right.\nonumber\\
&\left.\left. - t \left(r_\vartheta^2 + \Delta h_c^2\right)^{-\frac{\alpha _c}{2}}\right) \right] .
\end{aligned}
\end{equation}

\textit{\textbf{Proposition 4 (Laplace Transform of Aggregated Sensing Interference Power $I_{ar}$):}} The Laplace transform of the aggregated sensing interference power, conditioned on the distances between the typical ST and master BS as well as slave BS $j$, i.e., $R_0$ and $R_j$, is given by
\begin{equation}
  \begin{aligned}
    \mathcal {LT}_{I_{ar} |R_0, R_j }(t) = e^{\lambda_B \int_{0}^{2\pi} \mathcal{I}_1^{ri}(\vartheta) \,d\vartheta } ,
  \end{aligned}
\end{equation}
where
\begin{equation}
  \begin{aligned}
\mathcal{I}_1^{ri}(\vartheta) &= - \frac{r_\vartheta^{2 - \alpha_c} }{N_l(\alpha _c - 2)} \ _2F_1\left(1, 1 - \frac{2}{\alpha _c}; 2 - \frac{2}{\alpha _c}; - \frac{t}{N_l r_\vartheta^{\alpha _c} } \right)  .
\end{aligned}
\end{equation}

\section{Communication Performance Analysis}

\subsection{Communication Performance Metrics}
Building upon the analytical framework established in \cite{jiang2025network}, the following metrics are defined for evaluating communication performance.

\subsubsection{Per-CU metrics} To assess the reliability of individual communication links, the CCP for an arbitrary CU on a per-RE basis is defined as
\begin{equation}
\mathcal{P}_{uc}(T_c) \triangleq  \mathbb{P} \{\gamma _c > T_c\} ,
\end{equation}
where $T_c$ denotes the minimum SIR threshold required for successful coverage. 
The spectral efficiency of an arbitrary communication link is evaluated using the per-RE EE for a typical CU as
\begin{equation}
  \mathcal{R}_{uc}^e \triangleq \mathbb{E} \{\log (1 + \gamma _{c}) \} .
\end{equation}
Furthermore, the effective EE under multi-carrier aggregation over a single symbol duration (defined as aggregated EE) is given by
\begin{equation}
\mathcal{R}_{uc}^a \triangleq \frac{N}{\eta_cN_tN_l} \mathcal{R}_{uc}^e .
\end{equation}

\subsubsection{Per-BS metrics}
Given that $\eta_cN_tN_l$ CUs are simultaneously served by $N_l$ slave BSs, the per-BS performance metrics is defined as
\begin{subequations}
\begin{align}
&  \mathcal{P}_{bc}(T_c) \triangleq \eta_cN_t \mathbb{P} \{\gamma _c > T_c\} , \\
&  \mathcal{R}_{bc}^e \triangleq \eta_cN_t\mathbb{E} \{\log (1 + \gamma _{c}) \} , \\
&  \mathcal{R}_{bc}^a \triangleq \frac{N}{N_l}\mathbb{E} \{\log (1 + \gamma _{c}) \},
\end{align}
\end{subequations}
where $\mathcal{P}_{bc}$ denotes the average number of covered CUs per BS, $\mathcal{R}_{bc}^e$ represents the sum spectral efficiency per BS, and $\mathcal{R}_{bc}^a$ refers to the effective EE per BS under multi-carrier aggregation over a single symbol duration. Additionally, area-based performance (e.g., per square kilometer) can be obtained by multiplying the per-BS metrics by the BS density $\lambda_B$.

\subsection{Communication Performance Analysis}
Based on the signal power distributions derived in Section III, Per-RE CCP for a typical CU is given by the following theorem as proved in Appendix J in the supplemental materials.

\textit{\textbf{Theorem 1 (Per-RE CCP for a Typical CU in Heterogeneous Air-Ground ISAC CoMP Networks):}} In the considered network, the downlink per-RE CCP for a typical CU is given by
\begin{equation}
  \begin{aligned}
  \mathcal P_{uc}(T_c) &= \int_{0}^{r_c} \mathcal{P}_{uc | R_0}(T_c, r_0) f_{R_0}(r_0) \,d r_0,
\end{aligned}
\end{equation}
where the conditional CCP is expressed as
\begin{equation}
  \begin{aligned}
&\mathcal{P}_{uc | R_0}(T_c, r_0) = \\
&2 \int_{0}^{\infty} \Re \left\{\mathcal {LT}_{I_{ac} |R_0 }(2\pi \jmath s) \frac{\mathcal {LT}_{S_{ac} |R_0 } \left(-\frac{2\pi \jmath s}{T_c} \right) - 1}{2\pi \jmath s}\right\} \,ds .
\end{aligned}
\end{equation}

The per-RE EE for a typical CU is given by the following theorem as proved in Appendix K in the supplemental materials.

\textit{\textbf{Theorem 2 (Per-RE EE for a Typical CU in Heterogeneous Air-Ground ISAC CoMP Networks):}} In the considered network, the downlink per-RE EE for any CU is given by
\begin{equation}
  \begin{aligned}
    \mathcal{R}_{uc}^e
    = \int_{0}^{r_c} \mathcal{R}_{uc | R_0}^e(r_0) f_{R_0}(r_0) \,d r_0 ,
\end{aligned}
\end{equation}
where the conditional EE is
\begin{equation}
\mathcal{R}_{uc | R_0}^e(r_0) = \int_0^{\infty} \frac{1}{t} \left[1 - \mathcal {LT}_{S_{ac} |R_0 }(t)\right] \mathcal {LT}_{I_{ac} |R_0 }(t) \,dt .
\end{equation}

\begin{figure}[t]
  \centering
  \subfigure[3D view]{
  \label{Sensing Model1}
  \includegraphics[scale=0.5]{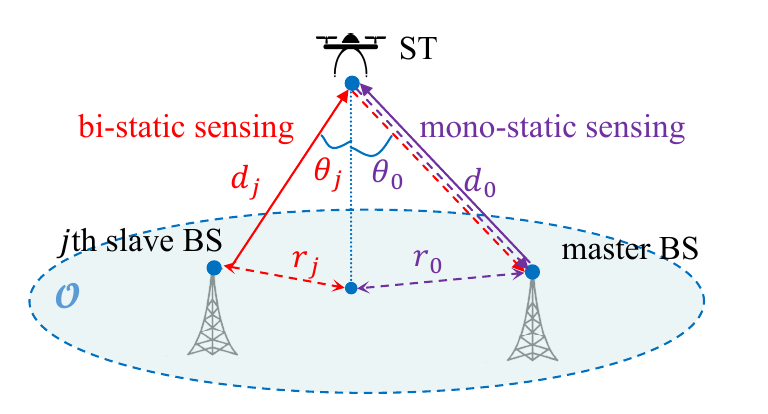}}
  \subfigure[Top-down view]{
  \label{Sensing Model2}
  \includegraphics[scale=0.5]{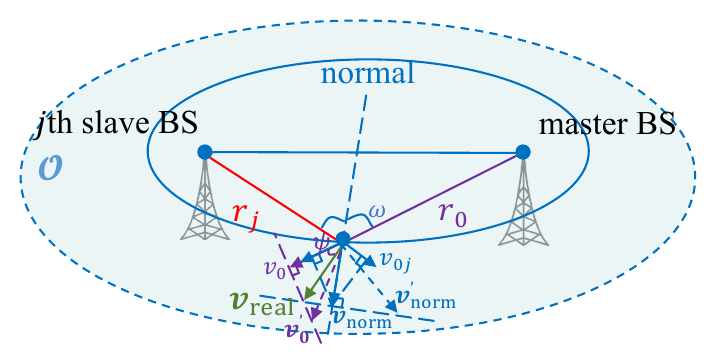}}
  \caption{Sensing network structure with low-altitude aerial STs.}
  \label{Sensing Model}
\end{figure}

\section{Sensing Performance Analysis}
The periodogram accurately reflects the strength of the reflected signal relative to the typical ST, ensuring that each link-specific estimated parameter is uniquely identifiable. The hybrid mono/bi-static sensing scheme is considered as illustrated in Fig. \ref{Sensing Model}, where the ST flies at a constant altitude $h_T$ with velocity $\bm v_{\text{real}} \in \mathbb{R}^2$. The master BS receives echo signals from both itself and cooperative BS $j$.\footnote{The momo-static and bi-static are implemented in a time-division manner.} The specific sensing process is introduced below.

\textit{a) Mono-Static Sensing:}
In mono-static sensing, the peak in the periodogram corresponding to the master BS-ST link provides estimate of the link distance $d_0 = \sqrt{r_0^2 + \Delta h_r^2}$ and the radial velocity. When the ST moves at velocity $\bm v_{\text{real}}$, the change in link distance is captured by the Doppler frequency $f_{D_0} = f_c \frac{ v_{\text{path}_0}}{c} = f_c \frac{2 \left\lVert \bm v_0\right\rVert \sin \theta_0}{c} $, where $v_{\text{path}_0}$ denotes the rate of path length change in the mono-static link, $\left\lVert \bm v_0\right\rVert  = \left\lVert \bm v_{\text{real}}\right\rVert \cos \psi$ is the horizontal radial speed, and $\left\lVert \bm v_0\right\rVert \sin \theta_0$ represents the radial speed (i.e., the projection of real velocity onto the link direction). 

\textit{b) Bi-Static Sensing:}
For the echo signal from bi-static cooperative BS $j$, the bi-static link distance is given by $d_{0j} = d_0 + d_j$. However, as far as velocity estimation is concerned, the situation is quite different. As shown in Fig. \ref{Sensing Model2}, for an ellipse with two BS as focuses, only the normal component of the velocity causes a change in the link distance. This change is captured by the Doppler frequency $f_{D_{0j}} = f_c \frac{ v_{\text{path}_{0j}}}{c} = f_c \frac{ \left\lVert \bm v_{0j}\right\rVert (\sin \theta_0 + \sin \theta_j)}{c}$, where $v_{\text{path}_{0j}}$ denotes the rate of path length change in the $j$-th bi-static link, $ \left\lVert \bm v_{0j}\right\rVert  = \left\lVert \bm v_{\text{norm}}\right\rVert \cos \omega $ is the horizontal radial speed, and $\left\lVert \bm v_{\text{norm}}\right\rVert = \left\lVert \bm v_{\text{real}}\right\rVert  \cos (\omega - \psi)$ denotes the horizontal normal speed.\footnote{Since the horizontal radial speeds of the master BS-ST and slave BS-ST links are both projections of the normal velocity along their respective directions, and the normal at the ST location bisects the angle between the two BSs (by the geometric property of ellipses), the two radial velocities are equal in magnitude, both equal to $\left\lVert \bm v_{0j}\right\rVert $.} 

\textit{c) Cooperative Hybrid Mono/Bi-static Sensing:}
Assuming parameter estimates are obtained from both mono-static and bi-static links, cooperative sensing can be performed. For instance, the link distance $d_j$ from the typical ST to slave BS $j$ can be derived. Similarly, the path-length change rates $v_{\text{path}_0}$ and $v_{\text{path}_{0j}}$ are estimated from Doppler frequencies. If the directions of the ST relative to each link are also estimated, the radial velocity $\bm v_0 $ and $\bm v_{\text{norm}} $ can be recovered. Although the radial velocity from any single link cannot uniquely identify the real ST velocity (e.g., in Fig. \ref{Sensing Model2}, $\bm v_0 $ and $\bm v_0^{'} $ yields the same radial velocity component, as $\bm v_{\text{norm}} $ and $\bm v_{\text{norm}}^{'} $), combining the estimates $\bm v_0 $ and $\bm v_{\text{norm}}$ allows unique identification of $\bm v_{\text{real}} $. This enables downstream sensing tasks such as localization and tracking.

\subsection{Sensing Performance Metrics}
This work assumes that the parameters for the mono-static sensing link are perfectly known, and focuses on deriving fundamental performance insights for bi-static sensing and cooperative hybrid mono/bi-static sensing. To evaluate the sensing performance, the sensing performance metrics are defined at both the link level and the network level.

\subsubsection{Link-level Metrics}
For the $j$-th bi-static sensing link, a false alarm occurs in a periodogram bin if the received interference power exceeds a threshold $\eta^{'}$, i.e., when $P_t \left(\frac{\lambda_c}{4\pi}\right) ^2 I_{ar} > \eta^{'}$. The CFAR per bin is thus defined as
\begin{equation}\label{Pbin}
  \mathcal{P}_{CFAR, bin} \triangleq \mathbb{P} \left\{P_t \left(\frac{\lambda_c}{4\pi}\right) ^2 I_{ar} > \eta^{'}\right\}  = \mathbb{P}\left\{ I_{ar} > \eta \right\}  .
\end{equation}
The overall CFAR over a CPI consisting of an $M\times N$ OFDM frame is given by
\begin{equation}\label{Pframe}
  \mathcal{P}_{CFAR, frame} =  1 - \left(1 - \mathcal{P}_{CFAR, bin}\right)^{NM}.
\end{equation}
Given a target frame-level CFAR $\mathcal{P}_{CFAR, frame}$, the threshold $\eta$ can be determined based on the distribution of $I_{ar}$. The corresponding SIR required for detection is $T_r = \eta/ \mathbb{E} \left\{I_{ar}\right\} $, where $\mathbb{E} \left\{I_{ar}\right\}$ is assumed known via pre-measurement and remains constant during sensing. The RDCP for the $j$-th bi-static sensing link, under a given CFAR constraint, is defined as
\begin{equation}
  \mathcal{P}_{0j}^{rcov}(T_r) \triangleq \mathbb{P} \{NM\gamma _{0j}^r > T_r \} ,
\end{equation}
which represents the probability that the typical ST is successfully detected by the $j$-th bi-static link, where the factor $NM$ accounts for the coherent processing gain of the periodogram \cite{sturm2009novel}. 

Calculating the CRB for the entire OFDM signal is highly complicated due to the unknown Doppler and delay parameters, which introduce random phases across the resource grid. Instead, the ACRB \cite{Braun14} is adopted to lower-bound the estimation errors for distance and velocity $\widehat{d_{0j}} $ and $\widehat{v_{0j}}$: 
\begin{subequations}
  \begin{align}
    \text{var}(\widehat{d_{0j}}) \geqslant \mathcal{ACRB}(\widehat{d_{0j}}) 
    &\triangleq \frac{6}{ (N^2 - 1)NM\gamma _{0j}^r } \left(\frac{c}{4\pi \Delta f} \right)^2  ,  \\
    \text{var}(\widehat{v_{0j}}) \geqslant \mathcal{ACRB}(\widehat{v_{0j}}) 
    &\triangleq \frac{6}{ (M^2 - 1)MN\gamma _{0j}^r } \left(\frac{c}{4\pi T f_c} \right)^2  .
  \end{align}
\end{subequations} 

\subsubsection{Network-level Metrics}
This work considers the $L \leq N_l$ nearest BSs to the typical ST for cooperative sensing.\footnote{Generally, the $L$ nearest BSs provide the highest average SIRs.} Since the estimated link distances $\left\{d_{0j}\right\}_{j = 1:L}$ are correlated due to common factor $d_0$, the corresponding peaks in the periodogram are statistically dependent. Therefore, joint detection across links is necessary, rather than treating the $L$ links independently. This approach avoids the accumulation of false alarms and leverages the inherent inter-link correlation. The RCDP for cooperative sensing is defined as 
\begin{equation}
  \begin{aligned}
    \mathcal{P}_{rcd}(T_r) \triangleq 1 - \prod\nolimits_{j=1}^{L} (1 - \mathcal{P}_{0j}^{rcov}( T_r)) ,
  \end{aligned}
\end{equation}
which represents the cumulative probability that at least one of the $L$ links successfully detects the ST. A higher RCDP indicates greater confidence in the presence of the ST. 

Although observations from multiple links can be fused for enhanced decision-making in cooperative sensing, this study focuses on revealing fundamental network-level performance. The design of specific multi-sensor fusion algorithms is left for future work.

Furthermore, as described in the cooperative sensing framework, the real velocity of the ST can be uniquely identified by combining the positions of any two BSs with their corresponding relative velocity estimates, via solving a system of linear equations. Accordingly, the ACRB for the actual speed of the ST can be expressed as a linear combination of the ACRBs of the individual links.

\subsection{Sensing Performance Analysis}
The Link-level RDCP and Link-level ACRB are given by the following theorems as proved in Appendix L and O in the supplemental materials.

\textit{\textbf{Theorem 3 (Link-level RDCP for a Typical ST in Heterogeneous Air-Ground ISAC CoMP Networks):}} The RDCP for the $j$-th bi-static sensing link is given by
\begin{equation}
  \begin{aligned}
    \mathcal{P}_{0j}^{rcov}( T_r) = \int_0^{r_c } \mathcal{P}_{0j}^{rcov | R_0}(T_r, r_0) f_{R_0}( r_0) \,d r_0  ,
\end{aligned}
\end{equation}
where the conditional RDCP is expressed as
\begin{equation}
  \begin{aligned}
    &\mathcal{P}_{0j}^{rcov | R_0}(T_r, r_0) = \\
    &\int_0^{r_0 + r_c } \mathcal {LT}_{I_{ar} |R_0, R_j }(m_r) f_{R_j}( r_j | r_0) \,d r_j ,
  \end{aligned}
\end{equation}  
with $m_r = 4\pi N_lT_r (d_0 d_j)^{\alpha _r}/(NM N_r\xi )$.

To establish the relationship between sensing SIR threshold $T_r$ and the required CFAR $\mathcal{P}_{CFAR, frame}$, two useful lemmas are introduced below as proved in Appendix M and N in the supplemental materials.

\textit{\textbf{Lemma 4.1 (The Explicit Expression between $T_r$ and $\mathcal{P}_{CFAR, frame}$ in Low BS Density Scenarios):}} Under low BS density, the relationship between sensing SIR threshold $T_r$ and the required CFAR $\mathcal{P}_{CFAR, frame}$ can be found by \textit{strongest interference approximation} (SIA) as
\begin{equation}
  \begin{aligned}
    T_r = - \ln \left[1 - (1 - \mathcal{P} _{CFAR,frame})^{\frac{1}{NM}} \right] .
  \end{aligned}
\end{equation}

\textit{\textbf{Lemma 4.2 (The Explicit Expression between $T_r$ and $\mathcal{P}_{CFAR, frame}$ in High BS Density Scenarios):}} Under high BS density, the relationship between sensing SIR threshold $T_r$ and the required CFAR $\mathcal{P}_{CFAR, frame}$ can be found as
\begin{equation}
  \begin{aligned}
    T_r = F_{I_{ar}}^{-1}\left(\left(1 - \mathcal{P}_{CFAR, frame}\right)^{\frac{1}{NM}} \right) /\mathbb{E} \left\{I_{ar}\right\} ,
  \end{aligned}
\end{equation}
where $F_{I_{ar}}(\cdot)$ denotes the CDF of $I_{ar}$ under a \textit{truncated-stable distribution} (TSD) approximation.

\textit{\textbf{Remark 3:}} Lemma 4.1 and 4.2 establish an explicit relationship between the sensing SIR threshold $T_r$ and $\mathcal{P}_{CFAR, frame}$. The SIA is more suitable for low BS density scenarios. In such sparse deployments, the aggregated interference is dominated by the few nearest, strongest interferers, resulting in a distribution with high kurtosis (peaked) and a heavy-tailed characteristic. In contrast, the TSD approximation is better suited for high BS density scenarios, where the interference from numerous, closely spaced sources is more uniform, leading to an aggregated distribution with lower kurtosis that is more symmetric and has thinner tails. In practice, the suitability of these distributions for modeling the actual aggregated sensing interference $I_{ar}$ can be evaluated using goodness-of-fit tests, such as the Kolmogorov-Smirnov statistic or the Kullback-Leibler divergence, thereby determining the applicable BS density regime for each model. Subsequently, Theorem 3 can be used to evaluate the RDCP under a given CFAR constraint.

\textit{\textbf{Theorem 4 (Link-level ACRB for a Typical ST in Heterogeneous Air-Ground ISAC CoMP Networks):}} The ACRBs for the estimated distance $\widehat{d_{0j}} $ and speed $\widehat{v_{0j}} $ in the $j$-th bi-static sensing link can be approximated by replacing the instantaneous SIR $\gamma _{0j}^r$ with its expectation:
\begin{subequations}
  \begin{align}
    \overline{\mathcal{ACRB}} \left(\widehat{d_{0j}}\right) 
    &= \frac{6}{ (N^2 - 1)NM \mathbb{E}\left\{\gamma _{0j}^r\right\} } \left(\frac{c}{4\pi \Delta f} \right)^2 \nonumber\\
    &= \frac{6}{ (N^2 - 1)\mathcal{I} _{0j}^r } \left(\frac{c}{4\pi \Delta f} \right)^2  , \\
    \overline{\mathcal{ACRB}} \left(\widehat{v_{0j}}\right) &=  \frac{6}{ (M^2 - 1)MN \mathbb{E}\left\{\gamma _{0j}^r\right\}} \left(\frac{c}{4\pi T f_c} \right)^2 \nonumber\\
    &= \frac{6}{ (M^2 - 1)\mathcal{I} _{0j}^r } \left(\frac{c}{4\pi T f_c} \right)^2 ,
  \end{align}
\end{subequations}
where $\mathcal{I} _{0j}^r = \int_{0}^{\infty }\mathcal{P}_{0j}^{rcov }(t) \,dt$.

\begin{table}
  \begin{center}
  \caption{Simulation Parameter Settings}
  \label{tab1}
  \resizebox{0.9\columnwidth}{!}{
  \begin{tabular}{| c | c | c |} 
  \hline
  Symbol & Value & Parameter Description\\
  \hline
  $N_t$ & 10 & Number of transmit antennas\\
  \hline
  $N_r$ & 10 & Number of receive antennas\\
  \hline
  $\lambda_B$ & $16 \ \mathrm{BSs/Km}^2$ & Density of BSs \\  
  \hline
  $N_l$ & 5 & Number of cooperative slave BSs within a single cluster \\
  \hline
  $r_c$ & $\sqrt{N_l/\pi\lambda_B}$ & Radius of cooperative cluster \\
  \hline
  $\eta_c$ & 0.5 & Loading factor of CU \\
  \hline
  $\eta_r$ & 1 - $\eta_c$ & Loading factor of ST \\
  \hline
  $h_B$ & $25 \ \mathrm{m}$ & Height of BS \\  
  \hline
  $h_U$ & $1.5 \ \mathrm{m}$ & Height of CU \\
  \hline
  $h_T$ & $100 \ \mathrm{m}$ & Height of ST \\
  \hline
  $f_c$ & $5.89 \ \mathrm{GHz}$ & Carrier frequency\\ 
  \hline
  $B$ & $10 \ \mathrm{MHz}$ & Bandwidth\\ 
  \hline
  $M$ & 50 & Total number of available symbols\\  
  \hline
  $N$ & 64 & Total Number of available subcarriers\\ 
  \hline
  $T$ & $8 \ \mathrm{\mu s}$ & Symbol duration\\ 
  \hline
  $\alpha _c$ & 4 & Pathloss exponent of the communication channel \\
  \hline
  $\alpha _r$ & 2 & Pathloss exponent of the sensing channel \\
  \hline
  $\xi $ & $1 \ \mathrm{m^2}$ & RCS of ST \\
  \hline 
  \end{tabular}}
  \end{center}
  \end{table}

\section{Simulation Results}
This section presents simulation results to evaluate the C\&S performance of the proposed cooperative air-ground ISAC CoMP network in an urban environment. The theoretical analyses derived in previous sections are validated through extensive \textit{Monte Carlo} (MC) simulations. Furthermore, the impacts of key system parameters on both C\&S performance are investigated. The default simulation parameters, unless otherwise specified, are configured according to the IEEE 802.11p standard \cite{nguyen2017delay} and summarized in Table \ref{tab1}.

\subsection{Communication Performance}

\begin{figure}[t]
  \centering
  \includegraphics[scale = 0.5]{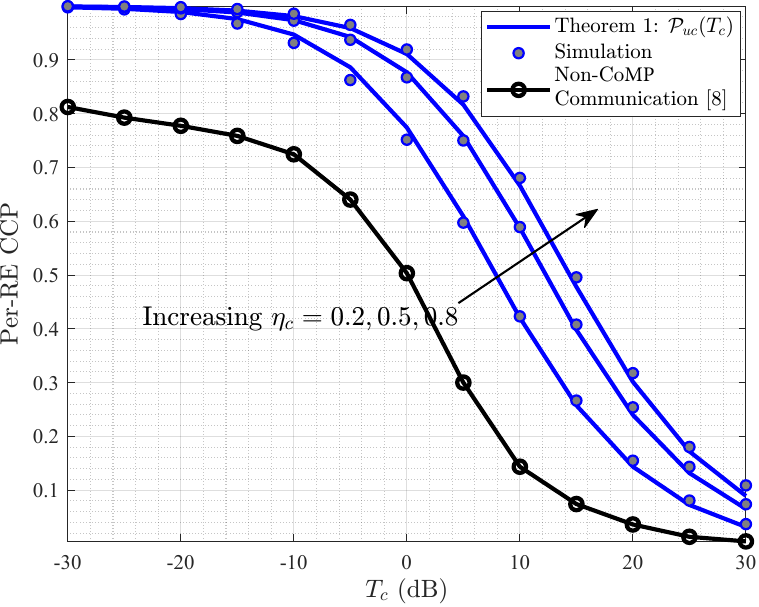}
  \caption{Per-RE CCP versus the communication SIR threshold $T_c$ under different loading factors $\eta_c \in \left\{0.2, 0.5, 0.8\right\} $.}
  \label{CCP}
\end{figure} 

Fig. \ref{CCP} illustrates the per-RE CCP for a typical CU versus the communication SIR threshold, $T_c$. The analytical results, derived from Theorem 1, are in close agreement with MC simulations across different communication loading factors, $\eta_c$, thereby validating the theoretical analysis. The figure also includes a performance comparison with a state-of-the-art non-CoMP communication baseline under the same $\lambda_B$ \cite{jiang2025network}. The results confirm that the proposed cooperative architecture yields superior CCP performance across all $T_c$ and $\eta_c$ values. This gain stems from the coherent gain of CoMP transmission and the effective mitigation of communication interference. Furthermore, it is observed that the CCP increases monotonically with $\eta_c$. This trend is consistent with the derivation of the per-BS BF gain $g_j^{cs}$ in Theorem 1, where $g_j^{cs}$ increases with $\eta_c$, leading to enhanced effective signal power and consequently improved SIR performance.

\begin{figure}[t]
  \centering
  \includegraphics[scale = 0.5]{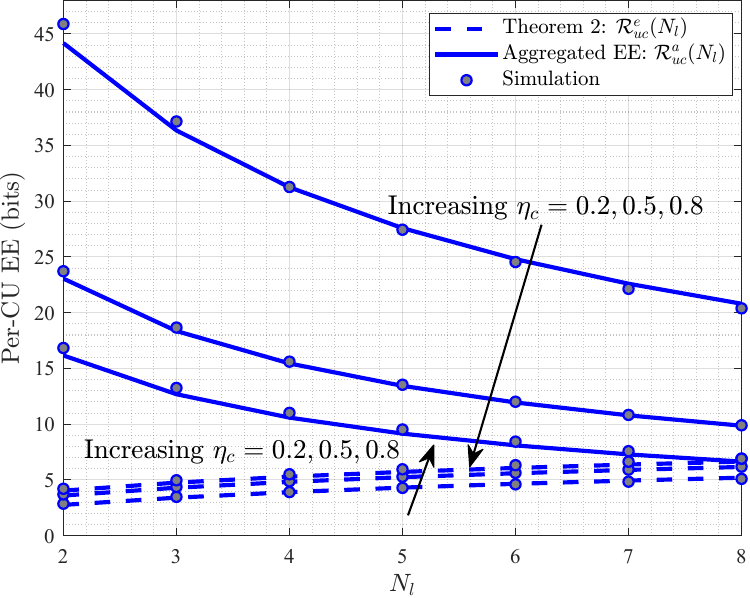}
  \caption{Per-CU EE as a function of the number of cooperative slave BSs per cluster $N_l$ under different loading factors $\eta_c \in \left\{0.2, 0.5, 0.8\right\}$.}
  \label{Per-CU EE Nl}
\end{figure} 

As illustrated in Fig. \ref{Per-CU EE Nl}, the per-CU EE is evaluated as a function of the number of cooperative slave BSs per cluster, i.e., $N_l$, under various loading factors $\eta_c$. Both per-RE and multi-carrier aggregated EE metrics are compared. The analytical results match well with MC simulations, confirming the accuracy of the theoretical analysis. The per-RE EE increases consistently with both $\eta_c$ and $N_l$, which can be attributed to three complementary mechanisms: (1) an enhancement in effective signal power due to higher per-BS BF gain at larger $\eta_c$, as established in Lemma 1; (2) a reduction in per-BS BF gain for the aggregated communication interference as $N_l$ increases; (3) improved coherent combining gain from a greater number of transmission sources (i.e., larger $N_l$). In contrast, the aggregated EE decreases with both $\eta_c$ and $N_l$. This is because although higher $\eta_c$ and $N_l$ improve individual link performance, the system's fixed total bandwidth must be shared among more CUs, leading to fewer subcarriers allocated per CU and consequently lower aggregated EE. This divergence underscores the critical balance between cooperative BF gains and spectral resource allocation in multi-cell networks.

\begin{figure}[t]
  \centering
  \includegraphics[scale = 0.5]{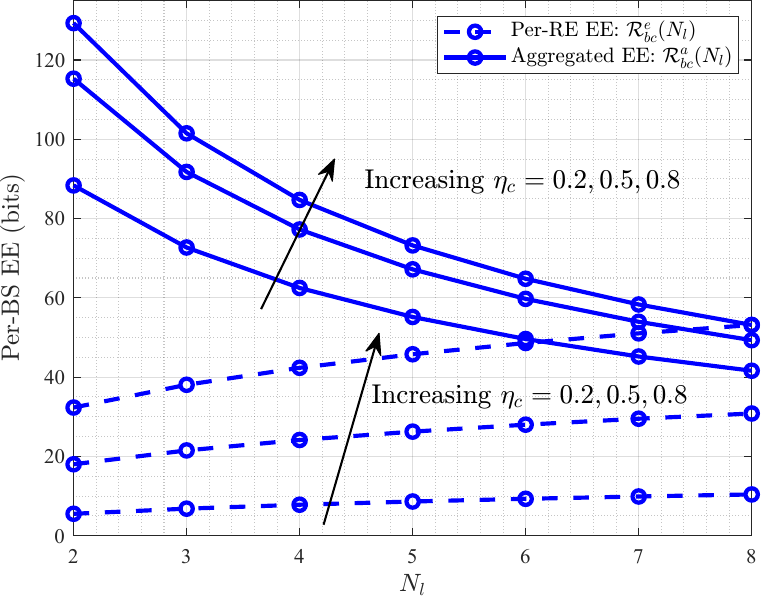}
  \caption{Per-BS EE as a function of the number of cooperative slave BSs per cluster $N_l$ under different loading factors $\eta_c \in \left\{0.2, 0.5, 0.8\right\} $.}
  \label{Per-BS EE Nl}
\end{figure}  

Fig. \ref{Per-BS EE Nl} plots the per-BS EE as a function of the number of cooperative slave BSs per cluster ($N_l$, ranging from 2 to 8), under different loading factors $\eta_c$. Different from the aggregated per-CU EE, the aggregated per-BS EE increases with higher $\eta_c$, since the subcarrier allocation per BS only depends on $N_l$ and is irrelevant with the loading factor $\eta_c$.

\begin{figure}[t]
  \centering
  \includegraphics[scale = 0.5]{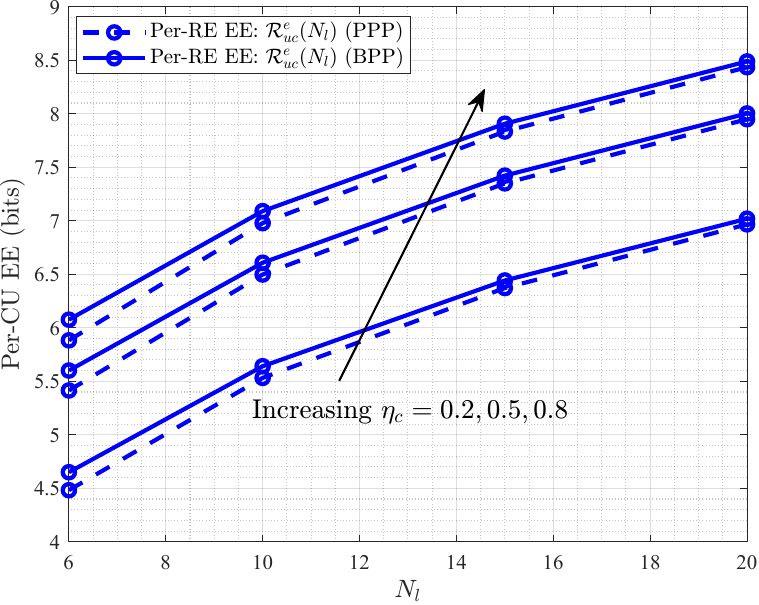}
  \caption{Comparison of Per-RE EE under BPP and PPP models.}
  \label{BPP vs PPP}
\end{figure} 

Fig. \ref{BPP vs PPP} further examines the asymptotic behavior of the per-RE EE under both BPP-based and PPP-based models, w.r.t the number of cooperative slave BSs per cluster ($N_l$, ranging from 6 to 20), evaluated across different loading factors $(\eta_c = 0.2, 0.5, 0.8)$. The results show that as $N_l$ increases, the BPP-based EE converges asymptotically to the corresponding PPP-based result, which is consistent with the theoretical analysis provided in Remark 1.

\begin{figure}[t]
  \centering
  \includegraphics[scale = 0.5]{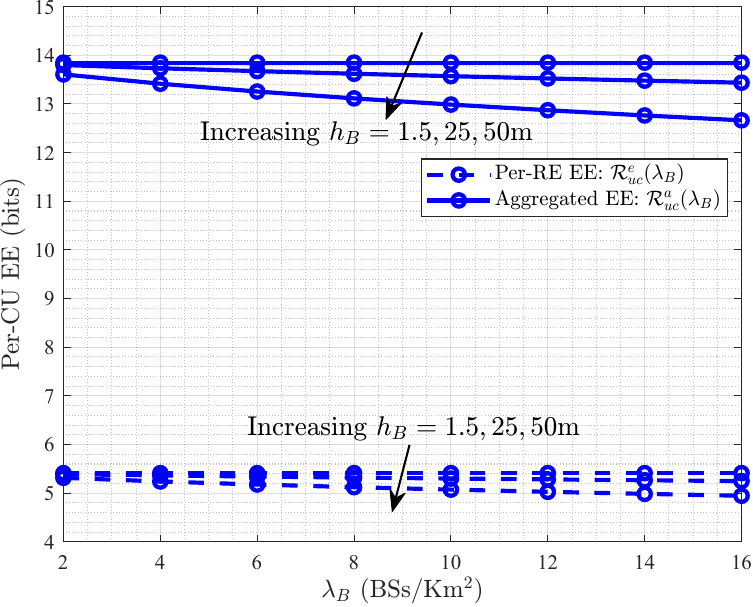}
  \caption{Per-CU EE as a function of the density of slave BSs $\lambda_B$ under different BS heights $h_B \in \left\{1.5, 25, 50\right\} $ m.}
  \label{Per-CU EE LambdaB}
\end{figure} 

Fig. \ref{Per-CU EE LambdaB} investigates the influence of BS density and height on the per-CU EE in terms of per-RE and aggregated EEs, respectively. As observed, both the per-RE EE and aggregated EE decrease with increasing BS density under the case of $\Delta h_c \neq 0$. This is because while the coherent combining gain from $N_l$ cooperative slave BSs improves due to the reduced service distances, the aggregated interference grows more significantly and ultimately dominates the performance. One can also observe that both the per-RE EE and aggregated EE decrease with increasing BS height. When the BS height is set to $h_B = 1.5 \ \mathrm{m}$ (i.e., $\Delta h_c = 0$), the scenario reduces to a conventional terrestrial network. In this case, the per-CU EE remains invariant with $\lambda_B$, since the increase in signal power is exactly counter-balanced by the increase in interference power.

\begin{figure}[t]
  \centering
  \includegraphics[scale = 0.5]{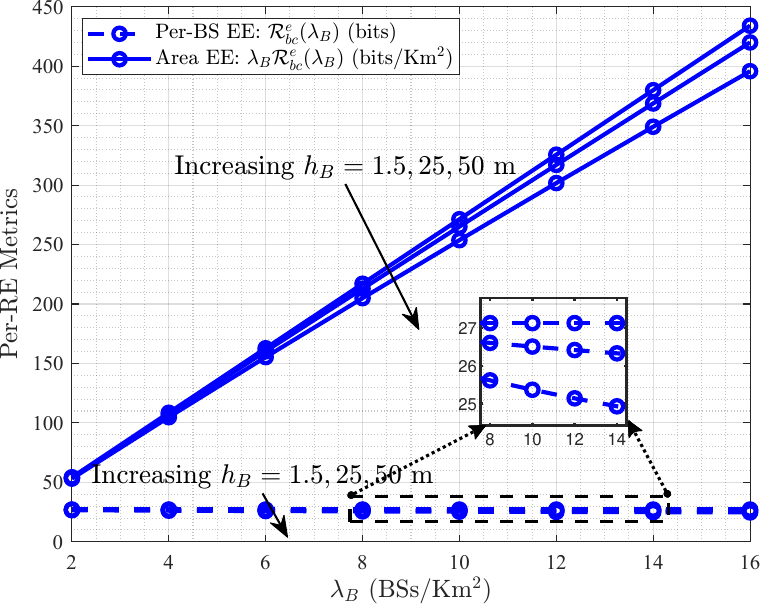}
  \caption{Per-RE EE as a function of the density of slave BSs $\lambda_B$ under different BS heights $h_B \in \left\{1.5, 25, 50\right\} $ m.}
  \label{Per-BS EE LambdaB}
\end{figure} 

As shown in Fig. \ref{Per-BS EE LambdaB}, the per-BS EE demonstrates a behavior consistent with the per-CU EE observed in Fig. \ref{Per-CU EE LambdaB}. However, the regional-level evaluation results indicate that area EE improves as BS density increases. This enhancement arises because the decline in per-CU link-level performance is compensated by a higher number of supported CUs per unit area. As a result, from a network-wide perspective, the area performance exhibits measurable improvement.

\subsection{Sensing Performance}
\begin{figure}[t]
  \centering
  \includegraphics[scale = 0.5]{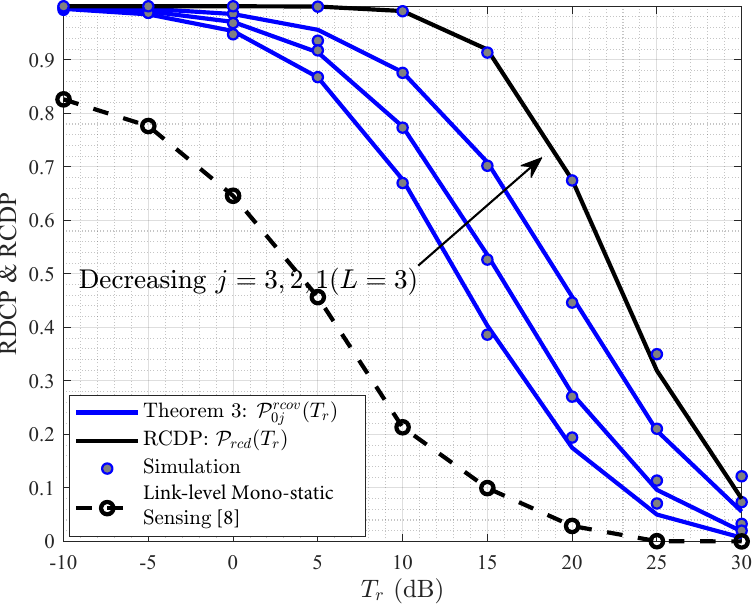}
  \caption{Link-level RDCP \& RCDP versus the sensing SIR threshold $T_r$ under different bi-static sensing links $j \in \left\{1, 2, 3\right\} $.}
  \label{RDCP}
\end{figure}  

Fig. \ref{RDCP} evaluates the link-level RDCP for multiple bi-static sensing links (indexed by $j = 1, 2, 3$), along with the RCDP for cooperative sensing. The simulation results align perfectly with the analytical results from Theorem 3. The performance comparison with a link-level mono-static sensing baseline \cite{jiang2025network} is also provided. The results demonstrate that the proposed cooperative sensing framework achieves significantly higher detection probabilities than the baseline across all $T_r$. This improvement quantitatively showcases the gain of cooperative sensing, which arises from the spatial diversity of distributed transmitters and the elimination of intra-cluster sensing interference. The analysis confirms that the RDCP decreases with the increasing $T_r$ and the longer propagation distance of the bi-static link. Moreover, owing to the cooperative sensing gain, the RCDP consistently surpasses the performance of each individual bi-static sensing link.

\begin{figure}[t]
  \centering
  \subfigure[Approximated ACRB of distance]{
  \label{ACRB d}
  \includegraphics[scale=0.5]{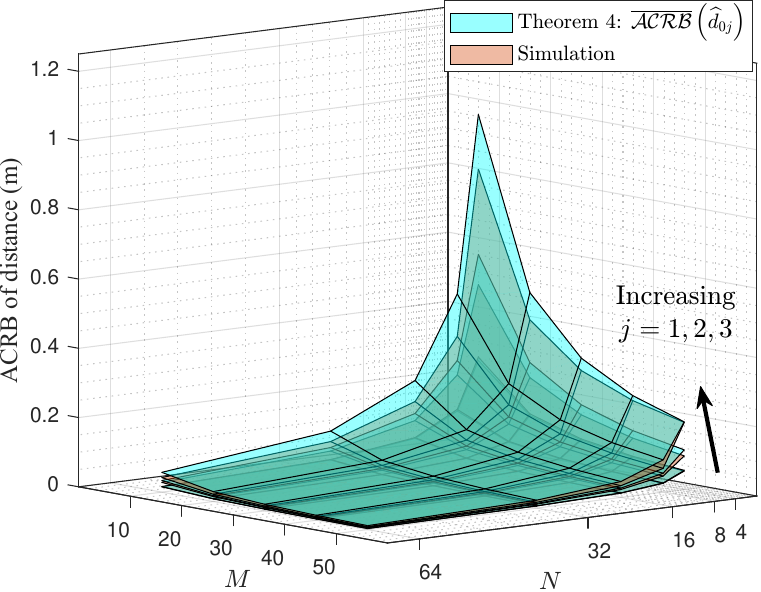}}
  \subfigure[Approximated ACRB of speed]{
  \label{ACRB v}
  \includegraphics[scale=0.5]{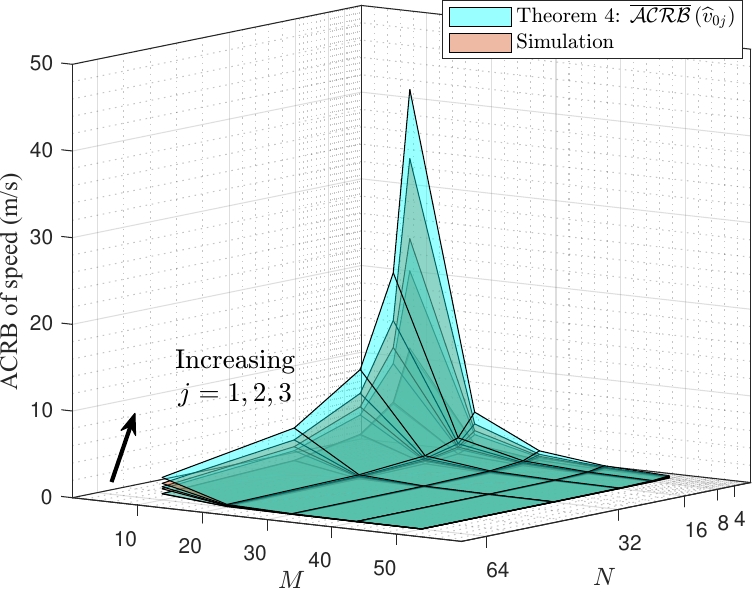}}
  \caption{The approximated ACRBs as a function of the numbers of OFDM subcarriers $N$ and symbols $M$ under different bi-static sensing links $j \in \left\{1, 2, 3\right\} $.}
  \label{Approximated ACRBs}
\end{figure}

Fig. \ref{Approximated ACRBs} illustrates the influence of the number of OFDM subcarriers $N$ and symbols $M$ on the approximated ACRBs for distance and speed estimation. The close alignment between simulation results and the analytical expressions derived in Theorem 4 confirms the validity of the theoretical model. It is observed that both distance and speed ACRBs decrease with increasing $N$ and $M$, which is consistent with the established principle in radar sensing that estimation accuracy improves with greater resource availability. Specifically, the ACRB for distance estimation is more sensitive to variations in $N$, exhibiting a steeper decline with additional subcarriers, while the ACRB for speed estimation is predominantly influenced by $M$.

\begin{figure}[t]
  \centering
  \includegraphics[scale = 0.5]{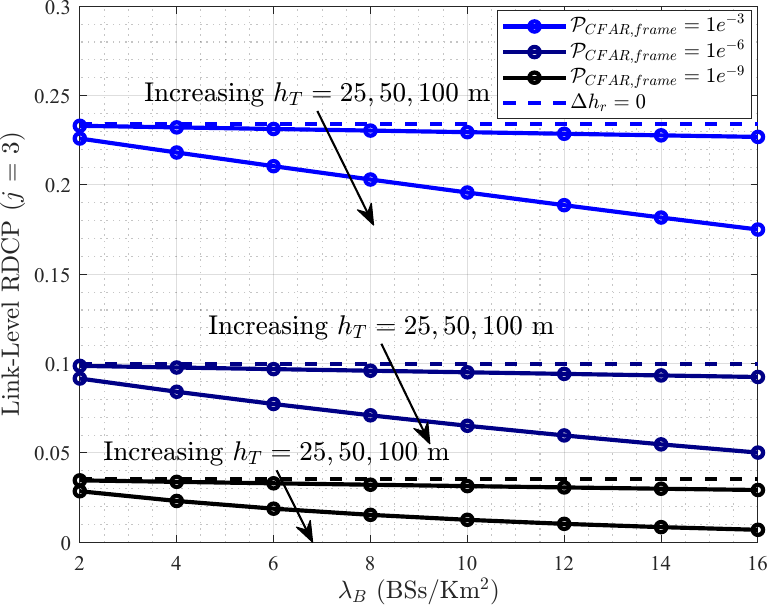}
  \caption{Link-level RDCP under given CFARs as a function of $\lambda_B$ under different ST heights $h_T \in \left\{25, 50, 100\right\}$.}
  \label{RDCP_lambdaB}
\end{figure}  

Fig. \ref{RDCP_lambdaB} depicts the influence of the BS density $\lambda_B$ and the ST height $h_T$ on the link-level RDCP under given CFAR. The results reveal two dominant trends: First, for elevated ST heights $(h_T = 50,100)$, the RDCP exhibits a clear decreasing trend with increasing $\lambda_B$, which is attributed to the accelerated growth of aggregated sensing interference that outweighs the increase in effective sensing signal. Second, in the case where $\Delta h_r =0$ (corresponding to a traditional terrestrial network, represented by the dashed curve), the RDCP remains invariant with $\lambda_B$, since the enhancement in signal power is exactly counterbalanced by a proportional increase in interference. Additionally, stricter CFAR requirements (i.e., lower $\mathcal{P}_{CFAR, frame}$) consistently lead to reduced RDCP, while lower ST altitudes generally improve sensing performance due to shorter propagation distance.

\section{Conclusion}
In this paper, the modeling and performance of a heterogeneous air-ground ISAC network based on CoMP architecture have been investigated for supporting the development of LAE. Moving beyond conventional terrestrial or single-cell ISAC systems, a two-tier network model incorporating both hexagonal and SG-based deployments has been introduced, along with a cooperative hybrid mono/bi-static sensing scheme. A comprehensive analytical framework has been developed to characterize the C\&S performance, accounting for critical channel and network effects. Simulation results underscore inherent trade-offs between C\&S performance, particularly under multi-BS cooperation and varying infrastructure density. These insights highlight the importance of balanced resource allocation and coordinated transmission strategies in practical multi-cell ISAC deployments. Our study provides fundamental insights and practical guidelines for the realization of scalable, efficient, and reliable ISAC networks in LAE scenarios. The SG-based analysis presented herein establishes essential \emph{average-performance benchmarks} for system design. Building upon this foundation, important future work includes the analysis of \emph{geometry-dependent performance fluctuations} (such as geometric dilution of precision and sensing blind zones in specific deployment snapshots) and their mitigation through dynamic resource optimization, to further enhance the reliability of ISAC networks for dynamic LAE scenarios.

\bibliographystyle{IEEEtran}
\bibliography{refs}

\appendix
\subsection{Proof of Corollary 1}
In the CoMP architecture, the typical CU receives coherent signals from multiple transmission points. Consequently, both the desired signal and interference are aggregated at the cluster level prior to power calculation. The analysis is complicated due to the distinct path losses incurred by signals originating from different slave BSs within the same cluster. To overcome this challenge, we derive approximate distributions for the effective signal power and the aggregated interference using Gamma 2nd-order moment matching \cite{forbes2011statistical}. Readers may refer to Section III of \cite{hosseini2016stochastic} for more details.

\subsection{Proof of Lemma 2}
Since $\bm w_l\in \mathbb{C} ^{N_tN_l \times 1}$ is a normalized isotropic vector on the complex unit sphere, $\mathbb{E}\left\{\bm w_l \bm w_l^H \right\} = \frac{1}{N_tN_l} \bm I_{N_tN_l} $. For a single slave BS, the subvector $\bm w_{lj} \overset{d}{= } \sqrt{\frac{1}{N_l}} \bm u \in \mathbb{C} ^{N_t \times 1}$, where $\bm u$ is isotropic on $\mathbb{C} ^{N_t \times 1}$, yielding $\mathbb{E}\left\{\bm w_{lj} \bm w_{lj}^H \right\} = \frac{1}{N_tN_l} \bm I_{N_t} $. Conditioned on the steering vector $\bm a_t^T(\theta_j)$, the conditional moments can be derived as
\begin{subequations}
  \begin{align}
    \rho _{| \bm a_t^T(\theta_j)} & := \mathbb{E}_{\bm w_{lj}} \left\{\left\lvert \bm a_t^T(\theta_j) \bm w_{lj} \right\rvert^2 \right\} \\
  &= \frac{1}{N_l} \mathbb{E}_{\bm u} \left\{\left\lvert \bm a_t^T(\theta_j) \bm u \right\rvert^2 \right\} \nonumber \\
  &\overset{(a)}{= } \frac{\bm a_t^H(\theta_j) \bm a_t(\theta_j)}{N_tN_l} , \nonumber\\
  \varrho _{| \bm a_t^T(\theta_j)} & := \text{Var}_{\bm w_{lj}} \left\{\left\lvert \bm a_t^T(\theta_j) \bm w_{lj} \right\rvert^2 \right\}  \\
  &= \frac{1}{N_l^2} \text{Var}_{\bm u} \left\{\left\lvert \bm a_t^T(\theta_j) \bm u \right\rvert^2 \right\} \nonumber \\
  &\overset{(b)}{= } \frac{N_t - 1}{N_l^2(N_t + 1)} \mathbb{E}^2_{\bm u} \left\{\left\lvert \bm a_t^T(\theta_j) \bm u \right\rvert^2 \right\} \nonumber\\
  &\overset{(c)}{= } \frac{N_t - 1}{N_t + 1} \rho ^2_{| \bm a_t^T(\theta_j)} \nonumber. 
\end{align}
\end{subequations}
where (a) follows from $\mathbb{E}_{\bm u} \left\{\left\lvert \bm a_t^T(\theta_j) \bm u \right\rvert^2 \right\} = \frac{\bm a_t^H(\theta_j) \bm a_t(\theta_j)}{N_t}$ and (b) comes from $\text{Var}_{\bm u} \left\{\left\lvert \bm a_t^T(\theta_j) \bm u \right\rvert^2 \right\} = \frac{N_t - 1}{N_t + 1} \mathbb{E}^2_{\bm u} \left\{\left\lvert \bm a_t^T(\theta_j) \bm u \right\rvert^2 \right\}$ \cite[Lemma 2]{jiang2025network}, and (c) comes from the second equation of (45a). 

\subsection{Proof of Corollary 2}
First, averaging over $\bm a_t^T(\theta_j)$ yields the unconditional moments:
\begin{subequations}
  \begin{align}
  \rho &:= \mathbb{E}_{\bm w_{lj}, \bm a_t^T(\theta_j)} \left\{\left\lvert \bm a_t^T(\theta_j) \bm w_{lj} \right\rvert^2 \right\} = \frac{1}{N_l} , \\       
  \varrho &:= \text{Var}_{\bm w_{lj}, \bm a_t^T(\theta_j)} \left\{\left\lvert \bm a_t^T(\theta_j) \bm w_{lj} \right\rvert^2 \right\} = \frac{N_t - 1}{(N_t + 1)N_l^2}. 
\end{align}
\end{subequations}
Next, applying Lemma 1 gives the shape and scale parameters as stated.

\subsection{Proof of Lemma 3}
The CDF and PDF can be derived by using a geometric argument analogous to that in \cite[Theorem 2.3.6]{mathai1999introduction}. Specifically, the CDF can be obtained by calculating the area of the intersection between the disk $\mathcal{D}(\bm p, r)$ and the disk $\mathcal{D}(\bm b_0, r_c)$, normalized by $\pi r_c^2$. This derivation considers two distinct cases, as illustrated in Fig. \ref{Distance Model}. Then, the PDF follows by differentiating the CDF w.r.t. $r$.  

\subsection{Proof of Corollary 3}
Note that $F_{R}(r_j|r_0) \triangleq p$ represents the probability that an arbitrary slave BS lies within the disk $\mathcal{D}(\bm p, r_j)$. Since the slave BS locations are conditionally i.i.d., the CDF of the $j$-th nearest distance is equivalent to the probability that at least $j$ BSs fall within $\mathcal{D}(\bm p, r_j)$. This corresponds to a well-known binomial probability problem \cite{wadsworth1960introduction} that can be solved as 
\begin{equation}
  \begin{aligned}
    F_{R_j}(r_j|r_0) = \mathbb{P} \{X \geqslant j \} = I_{p}( j, N_l - j + 1) 
  \end{aligned}
\end{equation}
where $X \sim B (N_l, p)$ and
\begin{equation}
  I_p(a, b) = \frac{\int_0^p t^{a-1}(1-t)^{b-1} \,dt}{\mathcal{B}(a,b)}
\end{equation}
denotes the regularized incomplete beta function. Further, the conditional PDF $f_{R_j}(r_j|r_0)$ can be derived by differentiating $F_{R_j}(r_j|r_0)$ w.r.t. $r_j$.

\subsection{Proof of Proposition 1}
The conditional Laplace transform of aggregated communication signal power under PPP model is defined as
\begin{equation}
  \begin{aligned}
    \mathcal {LT}_{S_{ac} |R_0 }(t) &\triangleq \mathbb{E}\left\{e^{-tS_{ac}}\right\} \\
    &= \mathbb{E}_{\Phi_S, g_j^{cs}} \left\{e^{ -t \sum _{j \in\Phi_S} d_j^{-\alpha _c}g_j^{cs}} \right\} \\
    &\overset{\text{(a)}}{=} \mathbb{E}_{\Phi_S} \left\{\prod\nolimits_{j\in \Phi_S} \mathbb{E}_{ g_j^{cs}} \left\{e^{-t d_j^{-\alpha _c} g_j^{cs}}\right\} \right\} \\ 
    &\overset{\text{(b)}}{=} \mathbb{E}_{\Phi_S} \left\{\prod\nolimits_{j\in \Phi_S} \underbrace{\left(1 + t d_j^{-\alpha _c}\right)^{-m_c} }_{\triangleq \mathcal{M} _{g_j^{cs}}(d_j) }\right\}  \\ 
    &\overset{\text{(c)}}{=} e^{\lambda_B \int_{0}^{2\pi} \underbrace{\int_0^{r_\vartheta } \left(\mathcal{M} _{g_j^{cs}}(r) - 1\right) r \,dr}_{\triangleq \mathcal{I}_1^{cs}(\vartheta) } \,d\vartheta } ,
  \end{aligned}
\end{equation}
where step (a) utilizes the independence between $\Phi_S$ and $g_j^{cs}$, step (b) applies the \textit{moment generating function} (MGF) of the Gamma distribution $X\sim \Gamma(\alpha , \beta )$ \cite{chiu2013stochastic}, i.e.,
\begin{equation}
  \begin{aligned}
\mathcal{M} _X(t) = \mathbb{E} _X \left\{e^{tX} \right\} = \left(1 - \beta t\right)^{-\alpha }  ,
  \end{aligned}
\end{equation}
and step (c) comes from the \textit{probability generating functional} (PGF) of the HPPP for the slave BSs \cite{haenggi2012stochastic}. 

Further, the above integral $\mathcal{I}_1^{cs}(\vartheta)$ can be processed as
\begin{equation}
  \begin{aligned}
    &\mathcal{I}_1^{cs}(\vartheta) = \int_0^{r_\vartheta } \left\{\left[1 + t (r^2 + \Delta h_c^2)^{- \frac{\alpha _c}{2}}\right] ^{-m_c} - 1\right\} r \,dr \overset{\text{(a)}}{=} \\
    &\frac{r_\vartheta^2 + \Delta h_c^2 }{2}\left[_2F_1\left(m_c, - \frac{2}{\alpha_c}; 1 - \frac{2}{\alpha_c}; - t \left(r_\vartheta^2 + \Delta h_c^2\right)^{- \frac{\alpha _c}{2}} \right) - 1 \right] \\
    & - \frac{\Delta h_c^2}{2} \left[_2F_1\left(m_c, - \frac{2}{\alpha_c}; 1 - \frac{2}{\alpha_c}; - t \Delta h_c^{-\alpha_c} \right) - 1 \right] ,
  \end{aligned}
\end{equation}
where (a) comes from the following integral identity
\begin{equation}
  \label{Gauss hyper1}
  \int \left(1 + ax^{-b}\right)^{-c} \! dx = x \ {_2F_1}\left(c, -\frac{1}{b}; 1-\frac{1}{b}; -ax^{-b} \right) .
\end{equation}

\subsection{Proof of Proposition 2}
The conditional Laplace transform of aggregated communication signal power under BPP model can be derived as
\begin{equation}
  \begin{aligned}
    &\mathcal {LT}_{S_{ac} |R_0 }(t) \triangleq \mathbb{E}\left\{e^{-tS_{ac}}\right\} \\
    &= \mathbb{E}_{\Phi_S} \left\{\prod\nolimits_{j\in \Phi_S} \left(1 + t d_j^{-\alpha _c}\right)^{-m_c} \right\}  \\ 
    &\overset{\text{(a)}}{=} \left\{\int_{0}^{r_c + r_0} \left[1 + t \left(r^2 + \Delta h_c^2 \right) ^{-\frac{\alpha _c}{2}}\right] ^{-m_c} f_R(r|r_0) \,dr \right\}^{N_l} \\ 
    &\overset{\text{(b)}}{=} \left\{\int_{0}^{r_c - r_0} \left[1 + t \left(r^2 + \Delta h_c^2 \right) ^{-\frac{\alpha _c}{2}}\right] ^{-m_c} \! \frac{2r}{r_c^2} \,dr + \int_{r_c - r_0}^{r_c + r_0} \right. \\
    &\left. \left[1 + t \left(r^2 + \Delta h_c^2 \right) ^{-\frac{\alpha _c}{2}}\right] ^{-m_c} \!\!\! \frac{2r}{\pi r_c^2}\arccos \left(\frac{r^2 + r_0^2 - r_c^2}{2r_0r}\right)  \! \,dr \right\}^{N_l} \\
    &\overset{\text{(c)}}{=} \left\{\frac{(r_c - r_0)^2 + \Delta h_c^2}{r_c^2} \right. \\
    &\cdot {_2F_1} \left(m_c, - \frac{2}{\alpha_c}; 1 - \frac{2}{\alpha_c}; - t \left[(r_c - r_0)^2 + \Delta h_c^2\right]^{- \frac{\alpha_c}{2}}\right)  \\
    & - \frac{\Delta h_c^2}{r_c^2} \ {_2F_1} \left(m_c, - \frac{2}{\alpha_c}; 1 - \frac{2}{\alpha_c}; - t \Delta h_c^{- \alpha_c}\right)  \\
    & + \int_{r_c - r_0}^{r_c + r_0} \left[1 + t \left(r^2 + \Delta h_c^2 \right) ^{-\frac{\alpha _c}{2}}\right] ^{-m_c} \frac{2r}{\pi r_c^2} \\
    &\left. \cdot \arccos \left(\frac{r^2 + r_0^2 - r_c^2}{2r_0r}\right) \,dr \right\}^{N_l}
  \end{aligned}
\end{equation}
where (a) follows from converting Cartesian to polar coordinates using density function of distance given by \eqref{unordered distance pdf} along with conditional i.i.d. property of $R$ with realization denoted by $r = \left\lVert \bm b_j - \bm u\right\rVert$, (b) comes from substituting the density function, $f_R(r|r_0)$, given by \eqref{unordered distance pdf}, and (c) comes from \eqref{Gauss hyper1}.

\subsection{Proof of Proposition 3}
The conditional Laplace transform of aggregated communication interference power is defined as
\begin{equation}
  \begin{aligned}
    &\mathcal {LT}_{I_{ac} |R_0 }(t) \triangleq \mathbb{E}\left\{e^{-tI_{ac}}\right\} = \mathbb{E}_{\Phi_I} \left\{\prod\nolimits_{j\in \Phi_I} \left(1 + t d_j^{-\alpha _c}\right)^{-\frac{1}{N_l}} \right\}  \\ 
    &= e^{\lambda_B \int_{0}^{2\pi} \underbrace{\int_{r_\vartheta }^\infty \left\{\left[1 + t \left(r^2 + \Delta h_c^2 \right) ^{-\frac{\alpha _c}{2}}\right]^{-\frac{1}{N_l}} - 1\right\} r \,dr}_{\triangleq \mathcal{I}_1^{ci}(\vartheta) } \,d\vartheta } .
  \end{aligned}
\end{equation}

Further, the above integral $\mathcal{I}_1^{ci}(\vartheta)$ can be processed as
\begin{equation}
  \begin{aligned}
\mathcal{I}_1^{ci}(\vartheta) &\triangleq \int_{r_\vartheta }^\infty \left\{\left[1 + t (r^2 + \Delta h_c^2)^{- \frac{\alpha _c}{2}}\right]^{-\frac{1}{N_l}} - 1\right\} r \,dr \\
&\overset{\text{(a)}}{=} \frac{r_\vartheta^2 + \Delta h_c^2}{ 2} \left[ 1 - {_2F_1}\left(\frac{1}{N_l}, - \frac{2}{\alpha _c}; 1 - \frac{2}{\alpha _c}; \right. \right. \\
&\left. \left. - t \left(r_\vartheta^2 + \Delta h_c^2\right)^{-\frac{\alpha _c}{2}}\right) \right] ,
  \end{aligned}
\end{equation}
where (a) follows from 
\begin{equation}
  \label{Gauss hyper2}
\int_0^u \frac{x^{\mu - 1}}{(1 + \beta x)^\nu} \,dx = \frac{u^{\mu}}{\mu} \ {_2F_1} \left(\nu, \mu; 1 + \mu; - \beta u\right).
\end{equation}

\subsection{Proof of Proposition 4}
The conditional Laplace transform of aggregated sensing interference power can be derived as
\begin{equation}
  \begin{aligned}
    &\mathcal {LT}_{I_{ar} |R_0, R_j }(t) \triangleq \mathbb{E}\left\{e^{-tI_{ar}}\right\} \\
    &\overset{\text{(a)}}{=} \mathbb{E}_{\Phi_I} \left\{\prod\nolimits_{j\in \Phi_I} \underbrace{\left(1 + \frac{t r_{0j}^{- \alpha_c }}{N_l} \right)^{-1} }_{\triangleq \mathcal{M} _{g_j^{ri}}(r_{0j}) }\right\}  \\ 
    &= e^{\lambda_B \int_{0}^{2\pi} \underbrace{\int_{r_\vartheta }^\infty \left(\mathcal{M} _{g_1^{ri}}(r) - 1\right) r \,dr}_{\triangleq \mathcal{I}_1^{ri}(\vartheta) } \,d\vartheta } ,
  \end{aligned}
\end{equation}
where step (a) applies the MGF of the exponential distribution $X\sim \text{Exp}(\lambda )$ \cite{chiu2013stochastic}, i.e.,
\begin{equation}
  \begin{aligned}
\mathcal{M} _X(t) = \mathbb{E} _X \left\{e^{tX} \right\} = \left(1 - \frac{t}{\lambda}\right)^{-1 } .
  \end{aligned}
\end{equation}

Further, the above integral $\mathcal{I}_1^{ri}(\vartheta)$ can be processed as
\begin{equation}
  \begin{aligned}
&\mathcal{I}_1^{ri}(\vartheta) \triangleq \int_{r_\vartheta }^\infty \left[\left(1 + \frac{t r^{- \alpha _c}}{N_l}\right) ^{-1} - 1\right] r \,dr \\ 
&\overset{\text{(a)}}{=} - \frac{r_\vartheta^{2 - \alpha_c} }{N_l(\alpha _c - 2)} \ _2F_1\left(1, 1 - \frac{2}{\alpha _c}; 2 - \frac{2}{\alpha _c}; - \frac{t}{N_l r_\vartheta^{\alpha _c} } \right) ,
  \end{aligned}
\end{equation}
where (a) comes from \eqref{Gauss hyper2}. 

\subsection{Proof of Theorem 1}
Conditioned on the distance $R_0 = r_0$ from the master BS to the typical CU, the per-RE conditional CCP is defined as
\begin{equation}
  \begin{aligned}
    &\mathcal{P}_{uc | R_0}(T_c, r_0) \triangleq  \mathbb{P}\left\{\gamma _{c} > T_c | R_0 = r_0 \right\} \\
    &= \mathbb{P}\left\{I_{ac} < \left. \frac{S_{ac}}{T_c} \right| R_0 = r_0 \right\} , \\
    &= \mathbb{E} _{S_{ac}} \left\{\int_{0}^{\frac{S_{ac}}{T_c}} f_{I_{ac}}(x) \,dx  \right\} \\
    &= \mathbb{E} _{S_{ac}} \left\{\int_{-\infty}^{\infty} f_{I_{ac}}(x) \mathbf{1}\left(0 \leq x \leq \frac{S_{ac}}{T_c} \right)  \,dx  \right\}   
  \end{aligned}
  \end{equation}
  \begin{equation}
    \begin{aligned}
    &\overset{\text{(a)}}{=} \mathbb{E} _{S_{ac}} \left\{\int_{-\infty}^{\infty} \mathcal {LT}_{I_{ac} |R_0 }(2\pi \jmath s) \frac{e^{2\pi \frac{S_{ac}}{T_c} \jmath s } - 1}{2\pi \jmath s}  \,ds \right\} \nonumber \\
    &\overset{\text{(b)}}{=} \int_{-\infty}^{\infty} \mathcal {LT}_{I_{ac} |R_0 }(2\pi \jmath s) \frac{\mathcal {LT}_{S_{ac} |R_0 } \left(-\frac{2\pi \jmath s}{T_c} \right) - 1}{2\pi \jmath s}  \,ds \nonumber \\
    &\overset{\text{(c)}}{=} 2 \int_{0}^{\infty} \Re \left\{\mathcal {LT}_{I_{ac} |R_0 }(2\pi \jmath s) \frac{\mathcal {LT}_{S_{ac} |R_0 } \left(-\frac{2\pi \jmath s}{T_c} \right) - 1}{2\pi \jmath s}\right\}  \,ds , \nonumber
  \end{aligned}
\end{equation}
where $f_{I_{ac}}(x)$ denotes the PDF of the aggregated communication interference power $I_{ac}$. Step (a) follows from the Plancheral-Parseval theorem, step (b) is derived using Fubini's theorem, while step (c) comes from the conjugate property of the Laplace transform.

\subsection{Proof of Theorem 2}
From \cite{hamdi2010useful}, for nonnegative, uncorrelated RVs $X$ and $Y$, the following identity holds
\begin{equation}
  \mathbb{E} \left\{\log \left(1 + \frac{X}{Y}\right) \right\} = \int_0^{\infty} \frac{1}{t} \left(1 - \mathbb{E}\left\{e^{-tX}\right\} \right) \mathbb{E}\left\{e^{-tY}\right\} \,dt .
\end{equation}

Conditioned on $R_0 = r_0$, and identifying $X = S_{ac}$, $Y = I_{ac}$, the conditional EE is
\begin{equation}
  \begin{aligned}
    &\mathcal{R}_{uc | R_0}^e(r_0)\triangleq \mathbb{E} \{\log _2(1 + \gamma _{c}) |R_0 = r_0\} \\
  &= \int_0^{\infty} \frac{1}{t} \left(1 - \mathbb{E}\left\{e^{-tS_{ac}}\right\} \right) \mathbb{E}\left\{e^{-tI_{ac}}\right\} \,dt \\
  &= \int_0^{\infty} \frac{1}{t} \left[1 - \mathcal {LT}_{S_{ac} |R_0 }(t)\right] \mathcal {LT}_{I_{ac} |R_0 }(t) \,dt .
\end{aligned}
\end{equation}

\subsection{Proof of Theorem 3}
Conditioned on the distance $R_0 = r_0$ from the typical ST to the master BS, the link-level conditional RDCP is derived as
\begin{equation}
  \begin{aligned}
    &\mathcal{P}_{0j}^{rcov | R_0}(T_r, r_0) = \\
    &\int_0^{r_0 + r_c } \mathcal{P}_{0j}^{rcov | R_0, R_j}(T_r, r_0, r_j) f_{R_j}( r_j | r_0) \,d r_j ,
  \end{aligned}
\end{equation} 
where the conditional RDCP is defined as
\begin{equation}
  \begin{aligned}
    &\mathcal{P}_{0j}^{rcov | R_0, R_j}(T_r, r_0, r_j) \triangleq \mathbb{P}\left\{ NM\gamma _{0j}^r > T_r | R_0 = r_0, R_j = r_j\right\} \\
    &= \mathbb{P} \left\{g_j^{rs} > \underbrace{\frac{4\pi T_r (d_0 d_j)^{\alpha _r}}{NM N_r\xi }}_{\triangleq q} I_{ar} | R_0 = r_0, R_j = r_j \right\} \\
    &\overset{(a)}{\approx} \mathbb{E}_{I_{ar}} \left\{e^{- N_l q I_{ar}} | R_0 = r_0, R_j = r_j \right\} \\
    &= \mathcal {LT}_{I_{ar} |R_0, R_j }(N_lq)  ,
  \end{aligned}
\end{equation} 
in which (a) comes from the CDF of an approximate exponential distribution of $g_j^{rs}$.  

\subsection{Proof of Lemma 4.1}
Under low BS density, the aggregated sensing interference can be approximated via SIA. The specific details can refer our previous work \cite[Proposition 3]{jiang2025network}.  

\subsection{Proof of Lemma 4.2}
In high BS density scenarios, the distribution of the aggregated sensing interference $I_{ar}$ can be approximated via a TSD. According to \cite{2025arXiv251003642J}, the \textit{characteristic function} (CF) of aggregated sensing interference $I_{ar}$ can be approximated by TSD as
\begin{equation}
  \begin{aligned}
    \label{TSD}
    \Upsilon _{I_{ar}}(\omega) = \text{exp}\left(c_{I_{ar}} \Gamma(-\alpha_{I_{ar}} )\left[(g_{I_{ar}} - \jmath \omega)^{\alpha_{I_{ar}}} - g_{I_{ar}}^{\alpha_{I_{ar}}} \right]\right)   , 
  \end{aligned}
\end{equation} 
where $\alpha_{I_{ar}} = 2/\alpha_c$ denotes the characteristic exponent, and the parameters $c_{I_{ar}}$ and $g_{I_{ar}}$ are defined as
\begin{subequations}
  \label{parameters of TSD}
  \begin{align}
    c_{I_{ar}} &= \frac{- \kappa _{I_{ar}} (1)}{\Gamma(-\alpha_{I_{ar}}) \alpha_{I_{ar}} \left[\frac{\kappa _{I_{ar}} (1)(1 - \alpha_{I_{ar}})}{\kappa _{I_{ar}} (2)} \right]^{\alpha_{I_{ar}} - 1} }  ,  \\
    g_{I_{ar}} &= \frac{\kappa _{I_{ar}} (1)(1 - \alpha_{I_{ar}}) }{\kappa _{I_{ar}} (2)} ,
  \end{align}
\end{subequations}
which represent the mirror stable distribution parameters and tempers the tail decay, respectively. 

Here, $\kappa_{I_{ar}}(n)$ denotes the $n$-th cumulant of the truncated aggregated sensing interference, which can be derived via Campbell's theorem \cite{haenggi2012stochastic}:
\begin{equation}
  \begin{aligned}
    \label{n-order cumulant of interference}
    \kappa_{I_v}(n) = \frac{2\pi \lambda_B}{n\alpha_c - 2}r_c^{2 - n\alpha_c} \mathbb{E} \{g_{ri}^n \} = \frac{2\pi \lambda_Br_c^{2 - n\alpha_c} \Gamma(1+n)}{n\alpha_c - 2} .
  \end{aligned}
\end{equation}

Further, the CDF of $I_{ar}$ can be obtained using the Gil-Pelaez's inversion theorem \cite{gil1951note}, i.e.,
\begin{equation}
  \begin{aligned}
    F_{I_{ar}}(x) &= \frac{1}{2} - \frac{1}{\pi} \int_{0}^{\infty } \frac{1}{\omega } \Im\left[\Upsilon _{I_{ar}}(-\omega) e^{\jmath\omega x}\right]  \,d\omega .
  \end{aligned}
\end{equation}

Thus, the CFAR per bin in the periodogram can be expressed as
\begin{equation}
  \begin{aligned}
  \mathcal{P} _{CFAR, bin} \triangleq \mathbb{P} \left\{I_{ar} > \eta \right\}= 1 - F_{I_{ar}}\left(\eta\right)  .
\end{aligned}
\end{equation}

Finally, by algebraic manipulation, we arrive at the final explicit expression between $T_r$ and $\mathcal{P}_{CFAR, frame}$.

\subsection{Proof of Theorem 4}
The expected value of the radar sensing SIR is given by
  \begin{align}
    &\mathbb{E} \left\{ \gamma _{0j}^r \right\}  
    \!= \! \int_{0}^{r_c} \! \int_{ 0}^{r_0 + r_c } \!\!\!\! \mathbb{E} \left\{\gamma _{0j}^r\left( r_0, r_j\right) \right\} \! f_{R_j}( r_j | r_0) \,d r_j f_{R_0}( r_0)  \,d r_0  \nonumber \\
    &= \int_{0}^{r_c} \int_{0}^{r_0 + r_c } \int_{t}^{\infty } \mathbb{P} \left\{\gamma _{0j}^r( r_0, r_j)>t | R_0 = r_0, R_j = r_j \right\} \,dt \nonumber \\
    &\cdot f_{R_j}( r_j | r_0) \,d r_j f_{R_0}( r_0) \,d r_0  \nonumber \\
    &\overset{\text{(a)}}{= } \frac{1}{NM}\underbrace{\int_{0}^{\infty } \mathcal{P}_{0j}^{rcov }(t) \,dt }_{\triangleq \mathcal{I} _{0j}^r}  ,
  \end{align}
where (a) comes from Theorem 3.  

\vspace{-18mm}
\begin{IEEEbiography}[{\includegraphics[width=1in,clip,keepaspectratio]{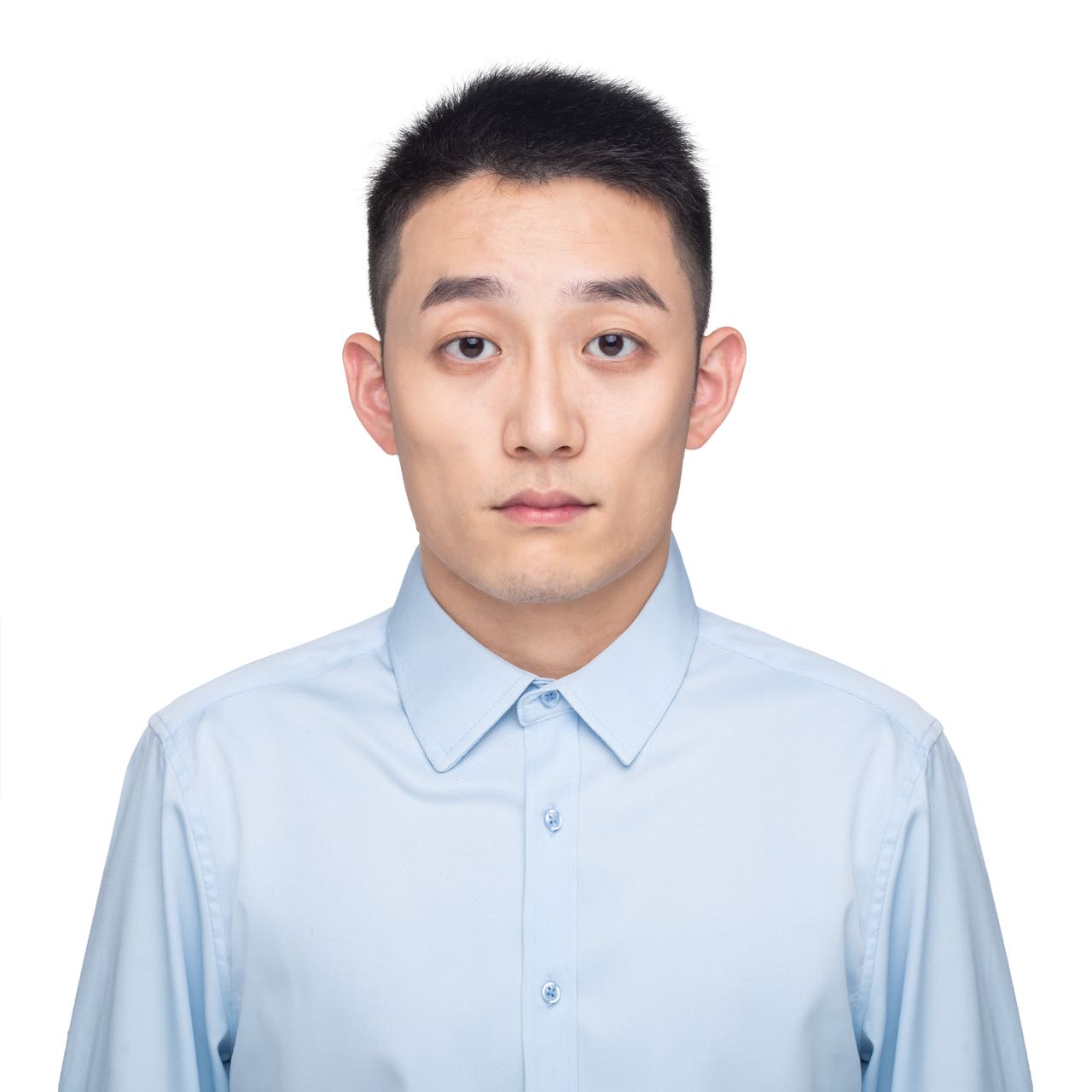}}]{Yihang Jiang}(Student Member, IEEE) received the M.Eng. degree from the Southern University of Science and Technology (SUSTech) in 2022. He is currently pursuing the Ph.D. degree with the School of Science and Engineering and Shenzhen Research Institute of Big Data, The Chinese University of Hong Kong-Shenzhen. His research interests include integrated sensing and communication, intelligent reflecting surfaces, and Wi-Fi sensing.
\end{IEEEbiography}

\vspace{-18mm}
\begin{IEEEbiography}[{\includegraphics[width=1in,clip,keepaspectratio]{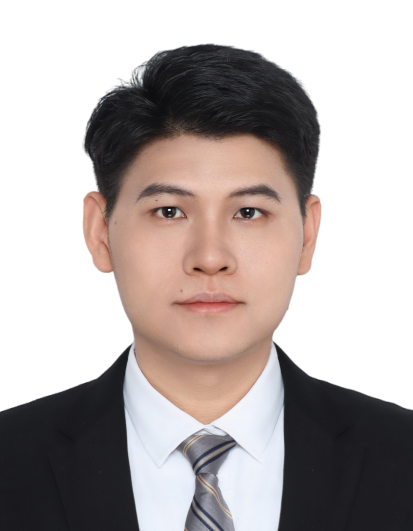}}]{Xiaoyang Li}(Member, IEEE) is currently an Assistant Professor with Southern University of Science and Technology. He received the Ph.D. degree from The University of Hong Kong. His research interests include integrated sensing-communication-computation and low-altitude wireless networks. He is a recipient of Young Elite Scientists Sponsorship Program by CAST, Forbes China 30 under 30, Young Elite of G20, Overseas Youth Talent in Guangdong, Overseas High-caliber Personnel in Shenzhen, Outstanding Research Fellow in Shenzhen, the Best Paper Award of IEEE 4th International Symposium on Joint Communications \& Sensing, the Exemplary Reviewers of IEEE Wireless Communications Letters and Journal of Information and Intelligence (JII). He has served as the Editor of JCIN and JII, and the Workshop Chairs of IEEE ICASSP/WCNC/PIMRC/Globecom/ICCC. 
\end{IEEEbiography}

\vspace{-18mm}
\begin{IEEEbiography}[{\includegraphics[width=1in,clip,keepaspectratio]{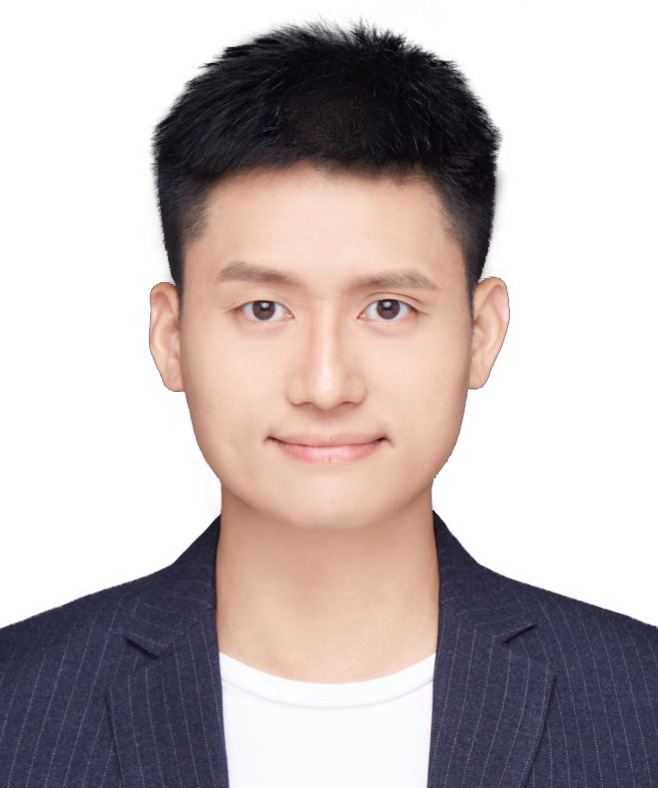}}]{Guangxu Zhu}(Member, IEEE) received the Ph.D. degree in electrical and electronic engineering from The University of Hong Kong in 2019. Currently he is a senior research scientist and deputy director of network and machine intelligence center at the Shenzhen research institute of big data, and an adjunct associate professor with the Chinese University of Hong Kong, Shenzhen. His recent research interests include edge intelligence, large foundation model, and integrated sensing and communication. He is a recipient of the 2023 IEEE ComSoc Asia-Pacific Best Young Researcher Award and Outstanding Paper Award, the World's Top 2\% Scientists by Stanford University, the “AI 2000 Most Influential Scholar Award Honorable Mention", the Young Scientist Award from UCOM 2023. He serves as associate editors at top-tier journals in IEEE, including IEEE TMC, TWC and WCL. He is the vice co-chair of the IEEE ComSoc Asia-Pacific Board Young Professionals Committee.
\end{IEEEbiography}

\vspace{-18mm}
\begin{IEEEbiography}[{\includegraphics[width=1in,clip,keepaspectratio]{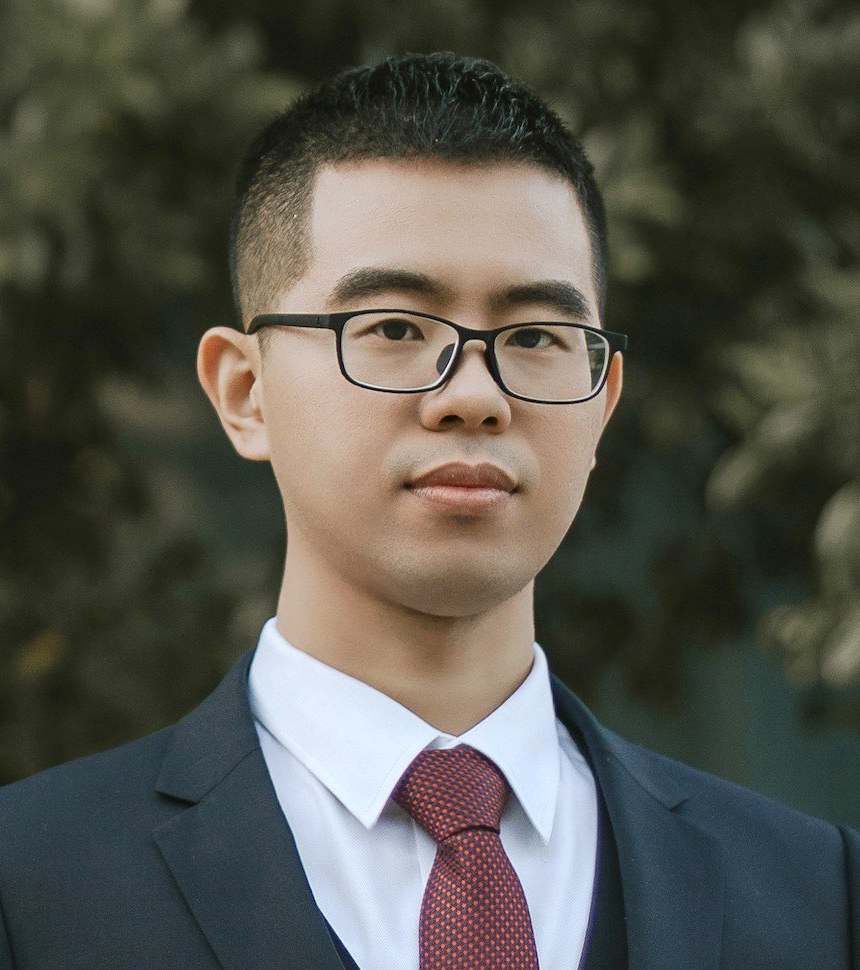}}]{Changsheng You} received the B.Eng. degree from the University of Science and Technology of China (USTC) in 2014 and the Ph.D. degree from The University of Hong Kong (HKU) in 2018. He is currently an Assistant Professor with the Southern University of Science and Technology, and was a Research Fellow with the National University of Singapore (NUS). His research interests include near-field communications, intelligent reflecting surfaces, UAV communications, edge computing, and learning. He currently serves an Editor of IEEE Transactions on Wireless Communications, IEEE Transactions on Mobile Computing, IEEE Transactions on Green Communications and Networking, IEEE Open Journal of the Communications Society, and IEEE Communications Letters. He received the IEEE ComSoc Leonard G. Abraham Prize in 2025, the IEEE ComSoc Best Tutorial Paper Award in 2023, the IEEE ComSoc Best Survey Paper Award in 2021, and the IEEE ComSoc Asia–Pacific Region Outstanding Paper Award in 2019. He is listed as Clarivate Highly Cited Researcher in 2024-2025 and IEEE ComSoc Asia-Pacific Best Yong Research in 2024.
\end{IEEEbiography}

\vspace{-18mm}
\begin{IEEEbiography}[{\includegraphics[width=1in,clip,keepaspectratio]{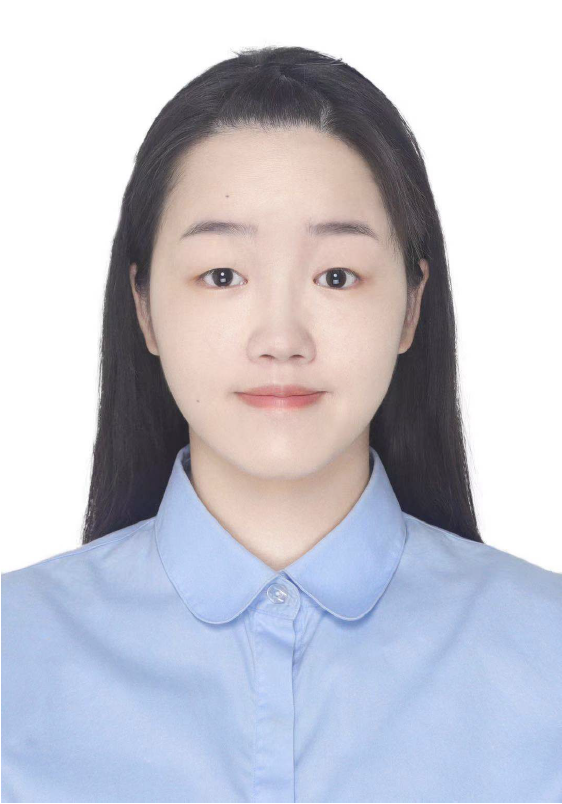}}]{Xiaowen Cao} (Member, IEEE) received the B.Eng. and Ph.D. degrees from the  Guangdong University of Technology in 2017 and 2022, respectively. She is now an assistant professor in College of Electronics and Information Engineering, Shenzhen University, Shenzhen, China, and a visiting scholar in Guangdong Provincial Key Laboratory of Future Networks of Intelligence, Shenzhen, China. Her research interests include edge learning, over-the-air computation, as well as integrated sensing, communication, and computation. She is a recipient of the World's Top 2\% Scientists by Stanford University, the Best Paper Award of IEEE JC\&S 2024, and the Exemplary Reviewer for IEEE WCL. She has served as a Co-Chair of  IEEE VTC-fall 2023 workshop, and a TPC Co-Chair of IEEE WCNC/PIMRC 2024 workshop. She is an associate editor of IEEE OJCOMS. 
\end{IEEEbiography}

\begin{IEEEbiography}[{\includegraphics[width=1in,clip,keepaspectratio]{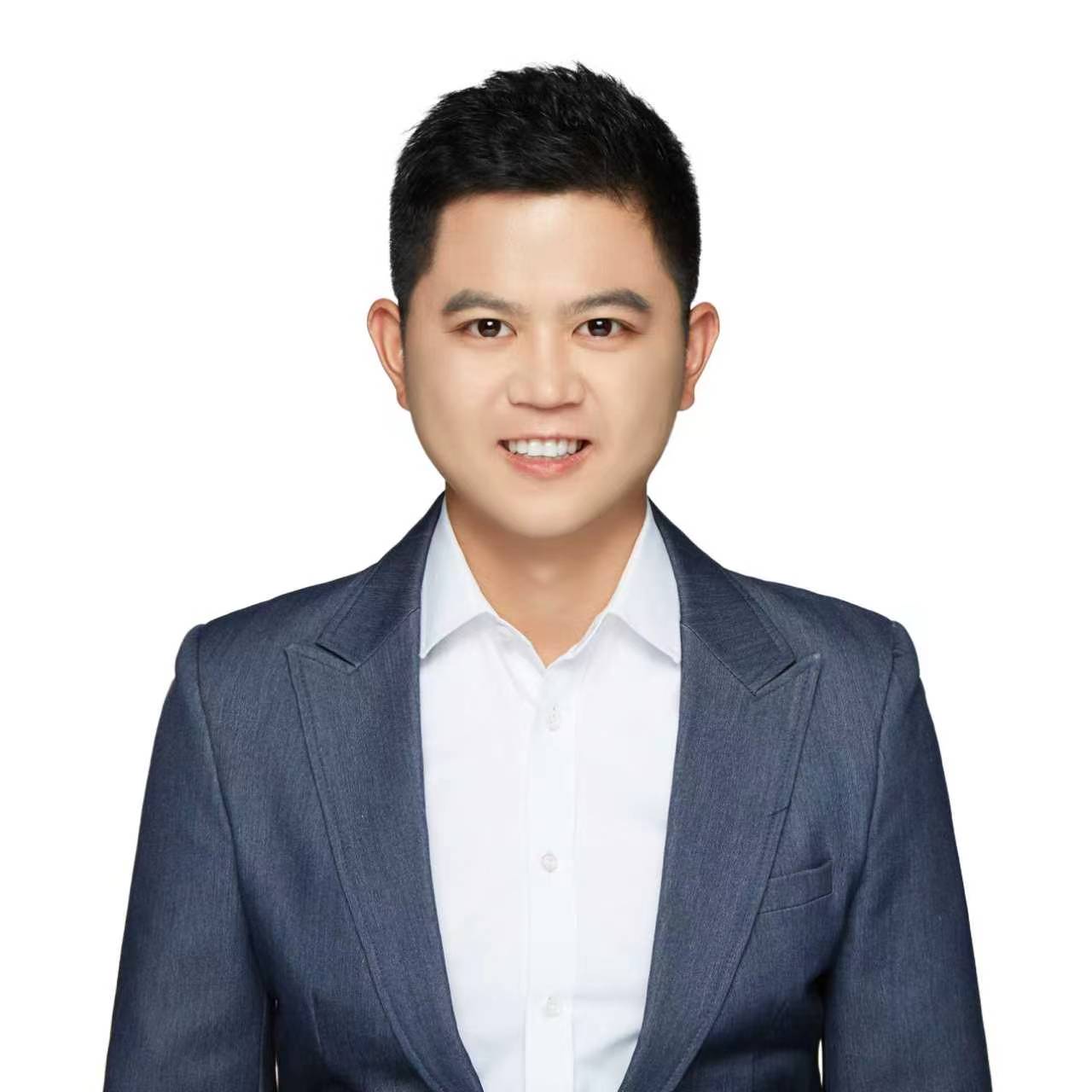}}]{Dingzhu Wen} (Member, IEEE) is an Assistant Professor at the School of Information Science and Technology in ShanghaiTech University. He received his Ph. D. degree from the Department of Electrical and Electronic Engineering of The University of Hong Kong in 2021. His research interests include brain-computer communication, edge AI, task-oriented communications, and integrated sensing-communication-computation. He has served as a co-organizer for workshops at flagship IEEE conferences including ICC, GlobeCom, WCNC, PIMRC, and VTC, and as a tutorial co-organizer at GlobeCom, WCNC, ICCC, and PIMRC. He has also chaired technical sessions at IEEE ICC, VTC, and WCSP. He was named as the Exemplary Reviewer for IEEE TCOM, TMC, and TNSE. He was awarded the IEEE GlobeCom 2023 Workshop Best Paper. He was selected as a World's Top 2\% Scientist (Stanford University \& Elsevier) and received the Excellent Mentor Award from ShanghaiTech University.
\end{IEEEbiography}

\vspace{-18mm}
\begin{IEEEbiography}[{\includegraphics[width=1in,clip,keepaspectratio]{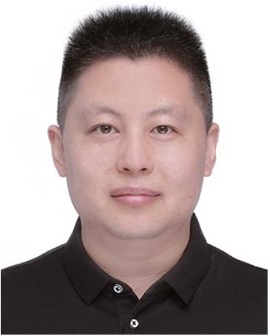}}]{Bingpeng Zhou} (Member, IEEE) received the Ph.D. degree from the Southwest Jiaotong University, Chengdu, China, in 2016. He was a Postdoctoral Fellow with the Hong Kong University of Science and Technology, Hong Kong, from 2016 to 2019. He was a Postdoctoral Researcher with Aalto University, Espoo, Finland, from 2019 to 2020. He was a Visiting Ph.D. Student with the 5G Innovation Centre, University of Surrey, Guildford, U.K., in 2015. He is currently an Associate Professor with the School of Electronics and Communication Engineering, Sun Yat-sen University, Shenzhen, China. He was selected for Major Talent Program of Guangdong Province for Distinguished Youth. His research interests include visible light-based positioning, integrated communication and sensing, Bayesian signal processing, and next-generation wireless networks.
\end{IEEEbiography}

\vspace{-18mm}
\begin{IEEEbiography}[{\includegraphics[width=1in,clip,keepaspectratio]{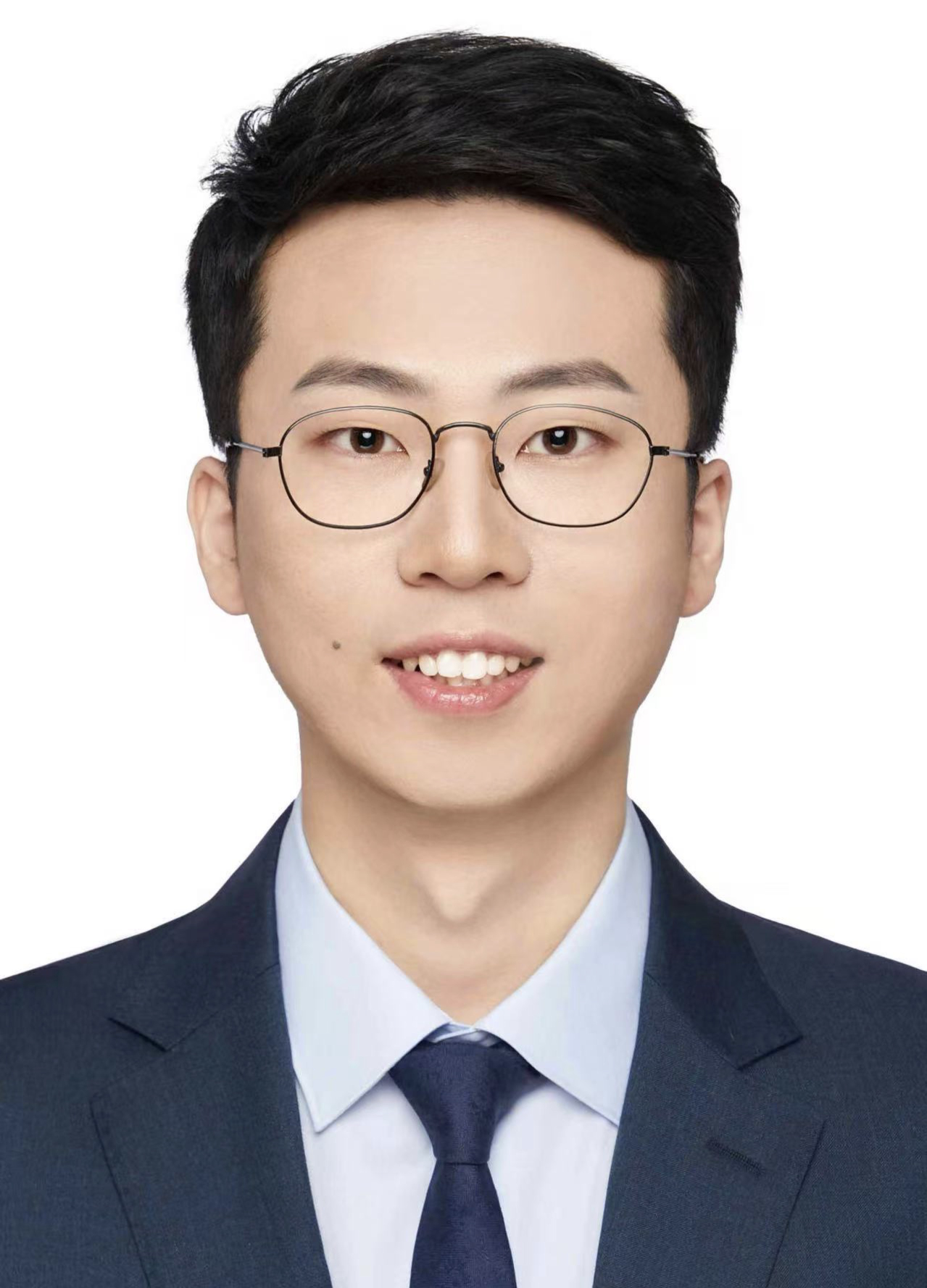}}]{Xinyi Wang} (Member, IEEE) received the B.Eng. and Ph.D. degrees in Information and Communication Engineering from Beijing Institute of Technology (BIT) in 2017 and 2022, respectively. From 2023 to 2024, he was a Postdoctoral Researcher at BIT, where he is currently an Associate Professor. He was a recipient of the Nomination Award for Outstanding Doctoral Dissertation by the China Education Society of Electronics (CESE). He has been recognized as an Exemplary Reviewer for the IEEE Transactions on Communications and IEEE Wireless Communications Letters. He was listed among the World’s Top 2\% Scientists by Stanford University for citation impact in 2025. His research interests include integrated sensing and communications, multi-carrier modulation techniques, and low-altitude wireless networks. He is a founding member of the IEEE ComSoc Special Interest Group (SIG) on LAWN, and has served as TPC members for multiple IEEE flagship conferences.
\end{IEEEbiography}

\vspace{-18mm}
\begin{IEEEbiography}[{\includegraphics[width=1in,clip,keepaspectratio]{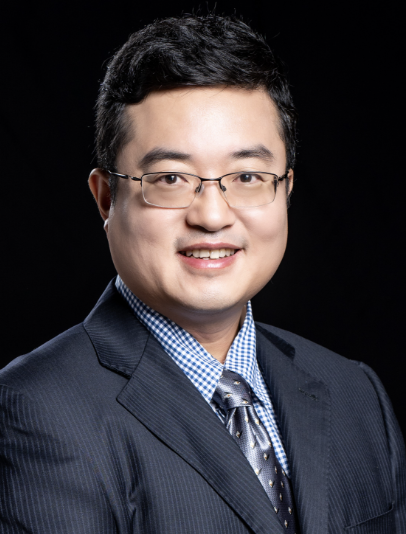}}]{Rui Zhang}(Fellow, IEEE) received the B.Eng. (first-class Hons.) and M.Eng. degrees from the National University of Singapore, Singapore, and the Ph.D. degree from the Stanford University, Stanford, CA, USA, all in electrical engineering. In 2010, he joined the Department of Electrical and Computer Engineering of National University of Singapore, where he is now a Provost’s Chair Professor. He has published over 300 journal papers and over 200 conference papers. He has published over 600 papers, all in the field of wireless communications and networks. He has been listed as a Highly Cited Researcher by Thomson Reuters/Clarivate Analytics since 2015. His current research interests include intelligent surfaces, reconfigurable antennas, radio mapping, non-terrestrial communications, wireless power transfer, AI and optimization methods. He received 18 IEEE Best Journal Paper Awards. He is a Fellow of the Academy of Engineering Singapore. \end{IEEEbiography}

\end{document}